\documentclass[namedreferences]{solarphysics}

\usepackage[hyperref,optionalrh]{spr-sola-addons} 
\usepackage{graphicx}        
\usepackage{color}           

\usepackage{hyperref,xcolor}
\hypersetup{
    colorlinks,
    linkcolor={red!60!black},
    citecolor={green!50!black},
    urlcolor={blue!80!black}
  }
\usepackage[all]{hypcap}

\usepackage[textwidth=6in,textheight=8in]{geometry}

\newcommand{\rc}[1]{{#1}} 

\newcommand{\rb}[1]{{#1}} 

\newcommand{\ra}[1]{{#1}} 

\chardef\us=`\_

\begin{document}

\begin{article}
\begin{opening}

\title{A stellar view of the Sun}

\author[addressref={aff1},email={karelschrijver@gmail.com}]{\inits{C.J.}\fnm{C.J.\ (Karel)}~\lnm{Schrijver}\orcid{0000-0002-6010-8182}}

\runningauthor{Karel Schrijver}
\runningtitle{A stellar view of the Sun}

\begin{abstract}
This invited memoir looks back on my scientific career that straddles the
solar and stellar branches of astrophysics, with 
sprinklings of historical context and personal opinion. Except
for a description of my life up to my Ph.D.\ phase, the structure is
thematic rather than purely chronological, focusing on those topics
that I worked on throughout substantial parts of my life: stars like
the Sun and the Sun-as-a-star, surface field evolution, coronal
structure and dynamics, heliophysics education, and space
weather. Luck and a broadly inquisitive frame of mind 
shaped a fortunate life on two continents, taking me from one amazing
mentor, colleague, and friend to another, working in stimulating
settings to interpret data from state-of-the-art space observatories.
\end{abstract}
\end{opening}

\section{Introduction}
\label{sec:introduction}

When Ed Cliver emailed me to talk ``re the memoirs that Solar Physics publishes
on a yearly basis,'' I expected that he wanted to discuss a slate of suitable
new candidates, which I could have readily provided. However, when, on
behalf of the nominating committee, he invited me to write the next memoir,
I could not believe my ears, even less so after he reminded me of the exceptional
colleagues who came before me, all of whom had worked in the field longer
than I have. Also, was I really ``in the field'' throughout my career?
After all, good parts of it were dedicated to the distant stars, the heliosphere,
and the impacts of solar activity on society. Then again, all those themes
are connected to the Sun -- past, present, and future. So, left humbled
by the invitation, I started to reconstruct my life.

The stars grabbed my interest and imagination from a very early age, the
Sun among their multitudes as the paragon of cool stars. It is not surprising
that almost all of my research into them has used observations made by
satellites: born in the year that NASA was founded, I grew up in the space
age, was just old enough to be transfixed by the Apollo Moon landings,
was in high school as Skylab's Apollo Telescope Mount was taking in the
Sun and as the first X-ray signals from cool stars were detected (starting
with Capella; \citet{1975ApJ...196L..47C} and
\citet{1975ApJ...202L..67M}, who, incidentally, worked at the two institutes that would
bracket my career) took on astrophysics as my major when the \textup{Einstein}
HEAO-2 observatory was launched, and started my Ph.D. research as \textup{EXOSAT}
and \textup{IUE} were being readied for launch to observe, among other things,
cool stars like the Sun. What \emph{is} surprising is that I have been improbably
fortunate to be able to work my dream job throughout my career, surrounded
and supported by wonderful colleagues, with --~as you will see in this
memoir~-- unbelievably fortuitous timing relative to the launch of a series
of space-based observatories.

Not only that, I have also been exceptionally fortunate in that my research
interests and projects were never limited by where I worked. Of course,
new projects would start in new settings, but I always had the opportunity
to continue, or at least gradually wrap up, projects as I found my footing
in a new job. Consequently, it proved easier to write this memoir guided
by research themes rather than by chronology, although it is not too far
off the latter. A~summary of when I worked where is nonetheless helpful
as a scaffold for my story, so you will find that in Section~\ref{sec:employment},
right after a description of my life up to my Ph.D. phase in Sections~\ref{sec:earlyyears}
and~\ref{sec:uu}. Then, follow the themes: stars like the Sun and the Sun-as-a-star
in Section~\ref{sec:sunstar}, the Sun's surface magnetic field in Section~\ref{sec:dispersal},
coronal structure and dynamics in Section~\ref{sec:flares}, heliophysics
education in Section~\ref{sec:heliophysics}, and space weather and its
impacts in Section~\ref{sec:swx}. Brief concluding Sections~\ref{sec:time}
and \ref{sec:lessons} wrap up this memoir.

I think that we are currently in a golden age of our research, with an
abundance of advanced observing platforms, powerful numerical laboratories,
and a diversity of technological tools that support us in our quest for
understanding. However, I also think that we are not optimally utilizing
these riches because the infrastructure that manages our research presents
several significant impediments to progress in our field. In what follows,
I give a number of examples regarding overcompartmentalization in research
and its funding, the relative weakness of the academic foundation of solar
physics, the piecemeal funding of researchers, and the high cost of space
missions. As these comments and their examples are spread throughout the
text and sometimes left implicit there, I summarize these points in the
final Section~\ref{sec:lessons}. There is also some advice to the next
generation sprinkled throughout the text and summarized in the final section.

An apology in advance: with each theme, I include a few references for
context to highlight earlier developments and also some of the more recent
work relevant to my activities, but please realize that this memoir is
not meant to serve as a review of any of its topics.

\section{Early Years}
\label{sec:earlyyears}

My parents were born into families of shopkeepers: sewing machines on my
father's side and, directly across the narrow village street of Baarn,
the Netherlands, a grocery shop on my mother's side. With only two years
difference between them, they told us how they played with each other and
with the other children in the quiet street, managed to sneak into the
attic of the Flora cinema right next door (where I, too, saw my first movies:
subtitled French-language comedies), and shared broken cookies from the
grocery that were unsuitable for sale (sometimes advancing that supply
by surreptitiously shaking the tins when my mother was left unsupervised
in the store). That untroubled village life ended in 1940, early in their
elementary school years, when the Second World War came to the Netherlands.
My parents' education continued only intermittently for the duration of
the war and they lost the opportunity to pursue higher education, which
was the main reason why they stimulated my younger sister and me to go
for it.

After the war, my parents lost sight of each other for some time as my
father went to work with a cousin in Sweden and was then drafted into the
army and stationed in New Guinea for almost two years. However, they found
each other again, married (without the blessing of their parents), and
rented a small apartment in a large villa where I was born two years later.
Among my earliest memories are those of the large estate where my grandmother
then worked: a home for elderly Indonesian expats, filled with mysterious
objects on the walls, a cage with tiny monkeys and another with colorful
birds, and above all the delicious smells perpetually emanating from the
large communal kitchen. The love of Indonesian food, already kindled by
my father's cooking of the dishes that he had encountered in New Guinea,
has always stayed with me.

When I was almost five years old, we moved into an apartment attached to
the factory where my father worked as an on-call repair technician. It
was located at the very edge of the small city of Soest: from the balcony
of our second-floor home we looked out over cow-filled pastures with the
seemingly endless forests of the Royal Domains in the distance, with dark
night skies above them providing views of the stars (in later years, the
flat roof provided a great location to mount my telescope). There were
only a few other homes nearby and hardly any children to play with, with
even those disappearing as their homes had to make way for a continuation
of the road in front of our apartment. As a result, I grew up largely entertaining
myself, something that I was never bothered by but that, I think, made
me into a rather shy, introverted child and young adult. Growing out of
that took decades.

If I had to characterize my main character trait in one word, it would
be curiosity. I~was building radios by age 11 (although the equations for
circuit design were well beyond me at that time), learned to install the
electrical circuits underneath the landscape for my M\"arklin model trains,
had a chemistry set to experiment with early in my teens, and spent hours
and hours reading all sorts of science books from the library. Then
there was the Apollo project: I followed flight after flight, often watching
it live on television deep in the night (as many of the main activities
were scheduled for daytime viewing in the USA).

Strangely, my wide-ranging curiosity nearly steered me away from my eventual
profession: despite good grades and an excellent score on the national
test in the final year, the head of my elementary school figured that I
would not succeed at higher education because, he argued, I had too many
interests. My parents accepted his counsel and, consequently, placed me
in a relatively low-level high school (MAVO). However, I was lucky: just
two years before that, a major reorganization in the Dutch education system
had been implemented. It allowed easy transition from one level of high
school to another. A~year later I had migrated two steps up to the highest
level, the atheneum. Instead of a 10-minute bicycle ride to school in Soest,
I was now fighting the all-too-frequent rain and wind on a 40-minute ride
to the nearest city, Amersfoort. I~loved physics and mathematics, thanks
to very inspiring teachers. Also, in retrospect, I should have been
most appreciative of my Dutch teacher: much to the puzzlement and amusement
of many of my classmates, he would bring in art books from his vast private
collection, hand around pictures of architecture, and play (mostly classical)
music on a rickety gramophone, but the result was that I still see literature,
art, music, and architecture in each other's context within an ever-changing
culture.

\begin{figure}
   \centerline{\includegraphics[width=0.5\textwidth,trim=0mm 100mm 0mm 70mm,clip=]{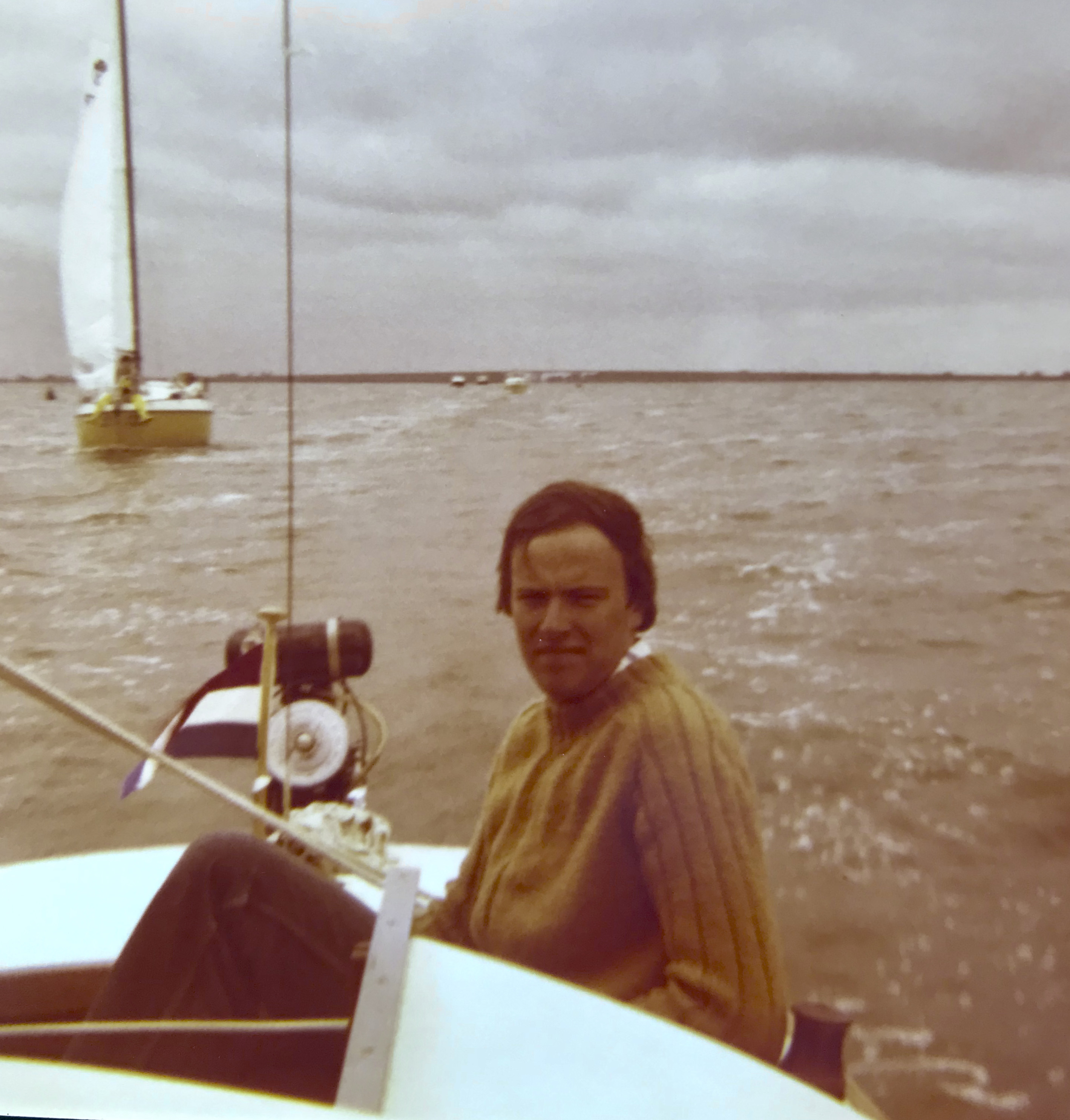}
              }
 \caption{Throughout my high-school years, my summer vacations were filled
with camping trips with my parents and sister, several weeks of factory
work, and a week or two of sailing through the northern province of Friesland
where small lakes (by American standards) are connected by a multitude
of waterways that bring one through the centers of picturesque towns.}
\label{fig:sail}
\end{figure}
Throughout all my school years, summers were times of camping, traveling,
and sailing (Figure~\ref{fig:sail}). At first, in my elementary school years,
we would camp on the coast of North Holland, where we could easily walk
into the dunes and onto the seemingly endless (and then still quiet) beaches.
There, my father would take the time to bring me up to speed where I had
lagged in my learning; we spent hours and hours working on multiplication
tables, often with ice cream as a promised reward. In later years, once
my parents had purchased a car, our vacations took us beyond the Dutch
borders: I remember campgrounds in the Ardennes in Belgium, on an island
in the Seine in western Paris, at Loch Ness in Scotland, in a pasture outside
London, at the base of a glacier in Switzerland {\ldots}
I documented many of these trips in super-8 home movies or, later, in slide
shows, learning how to accompany these with soundtracks using the earliest
electronic tools available to the general public.

Throughout these years, my father taught me to use my hands and not to
be afraid of mechanical or electrical jobs. During my Ph.D. phase, for
example, we once took apart the engine block of my 1100cc motorcycle to
find the cause of an odd sound that it had developed; it took us an entire
weekend before we had it back together but it ran smoothly ever after.
Also, he managed to convey some of his skills in 3D thinking: he eventually
became an instrument maker with an uncanny ability to reproduce or design
objects by using the variety of machines that surrounded his work area. I~still am grateful to him for his teachings whenever some job needs doing
around the house.

\section{Undergraduate Years}
\label{sec:uu}

Once I started attending the atheneum, the path to a university education
opened up. My father worked rotating shifts, sometimes even taking on double
duties, and often on Saturdays, to give my sister and me a chance for a
better education, but apart from my high school teachers, I knew no one
with a university degree at that time: no one in our extended family had
ever gone there. Fortunately, there was a system of financial support and
long-term, interest-free loans in place for students so I could pursue
my new dream. I~could not afford student housing, though, so I commuted
the 25 kilometers to Utrecht's Uithof campus and back every day, at first
on a moped, later on a 400cc motorcycle. I~enjoyed the ride over forest-covered
roads, although in the winters, when the Sun rose after classes started
and set before they ended, it was less enjoyable.

Physics and mathematics (my initial majors) were challenging but fun: classes
in the mornings and labs in the afternoons. I~particularly enjoyed some
of the unusual labs such as learning to develop film and to print photos
in evening astronomy labs (how quaint that seems nowadays), and simple
glass blowing. Also, I learned to play bridge in the lunch breaks with
my fellow students.

After the first two years, I flipped majors and minors, choosing to focus
more on astrophysics than on physics and mathematics. The way physics was
taught gave me the impression that most things were known with little left
to discover. In astronomy classes, I experienced exactly the opposite. I~never regretted that decision, although I have to admit to an unusual
degree of naivety back then: I gave hardly any thought to a career after
university. Things have fallen into place wonderfully, though, and I was
right at least about astronomy: there was, indeed, so much to discover!

In my third year, I elected to compute stellar-structure models in an elective
lab together with another enthusiastic student. To this day, I don't think
we deserved the grade that we were given; neither we nor the teaching assistant
who graded us understood the importance of the outer boundary condition
and the complexities involved in that. We approached advanced graduate
students (Paul Kuin and Piet Martens) for that challenge and they graciously
helped us toward the end goal. It taught me a lot about differential equations
and numerical methods, and a bit about computers, which, at that time would
require entire buildings for something less powerful than what we now have
in a laptop. We developed fragments of the code on teletypes: simple keyboards
connected to a mainframe, with instructions and responses typed out on
rolls of paper. Eventually, the full program was encoded on a stack of
punchcards which would be hand-carried to the building housing the large
IBM to be read in, and then hours later a printout would appear in one
of the many pigeonholes, arranged alphabetically by job name. Still, the
big physics computer was quite a bit more powerful than the Commodore 64
computer (with 64 kbytes of memory) that I bought and played with four
years later.

On the physics side, I emphasized plasma physics. That turned out to be
a wise choice given what the future had in store for me. The oral exam
(quite rare in Utrecht, and my first), in the vast office of Prof. Braams,
the director of the plasma physics institute in Nieuwegein, was
quite the ordeal. In mathematics, I favored the more applied side, in particular
numerical methods. I~took all the required classes in the astronomy curriculum,
including stellar evolution, degenerate matter, gravitation and cosmology,
astrophysical magnetohydrodynamics, etc., and, unavoidably, radiative transfer
that was taught by my future thesis adviser, Prof. Cornelis (Kees) Zwaan. I~still remember Kees' surprise in later years that his course was the
only one for which I did not take the exam, simply because it was unnecessary.
Coming to the very end of my studies, I had acquired more than enough points
to qualify for graduation, so why would I jeopardize my grade point average
on such a complex topic?

For what nowadays might be called my master's project, I chose to work
on the development of an X-ray position-sensitive proportional counter
at the Laboratory for Space Research (known by its Dutch abbreviation LRO)
at the other end of the city of Utrecht. The project developed into what
I have always liked in the years following --~combining observations with
modeling. In this case, it also involved learning that it is really hard
to straighten metal wires for a detection chamber from a roll and that
vibrations of vacuum pumps travel all over the place and can set up resonances
in tightly strung metal wires. It was fun, though, to see the green or
blue sparks (depending on whether methane or carbon dioxide was added to
the argon, xenon, or nitrogen gas) wherever the wires of the upper and
lower planes of the chamber came too close to each other. The results were
analyzed with LRO's PDP-11 computer that read out the measurements from
a paper tape, a very fragile carrier for punch holes that easily tore and
then would have to be repaired by adhesive tape, requiring a good memory
for encoding the eight bits of a character into zeros and ones, and a punch
to create the corresponding holes in the repaired tape. How the world has
changed in just four decades!

Discussions between the staff members brought me to the attention of Rolf
Mewe at LRO who, together with Kees Zwaan of the Astronomical Institute,
was looking for a student to work on X-ray spectroscopy of cool stars. I~don't think my experience with gas proportional counters had much to
do with their choice, although I did start that work on observations made
by the imaging proportional counter onboard the \textup{Einstein} HEAO-2 spacecraft.
It was not quite the direction that I was most interested in at the time
--~stellar evolution theory~-- but the work was to involve stars in different
evolutionary stages, so I accepted their offer for a three-month project.

Before formally graduating in March of 1981, I took a six-month course
to obtain the license qualifying me as a physics teacher at any of our
levels of high school --~one never knows, after all. It was very interesting
to learn about learning, then to actually teach three different age classes
in a Montessori high school in Zeist for a month, and in doing so gain
confidence in speaking before an (often restless) audience.

\begin{figure*}
   \centerline{\includegraphics[width=0.75\textwidth,clip=]{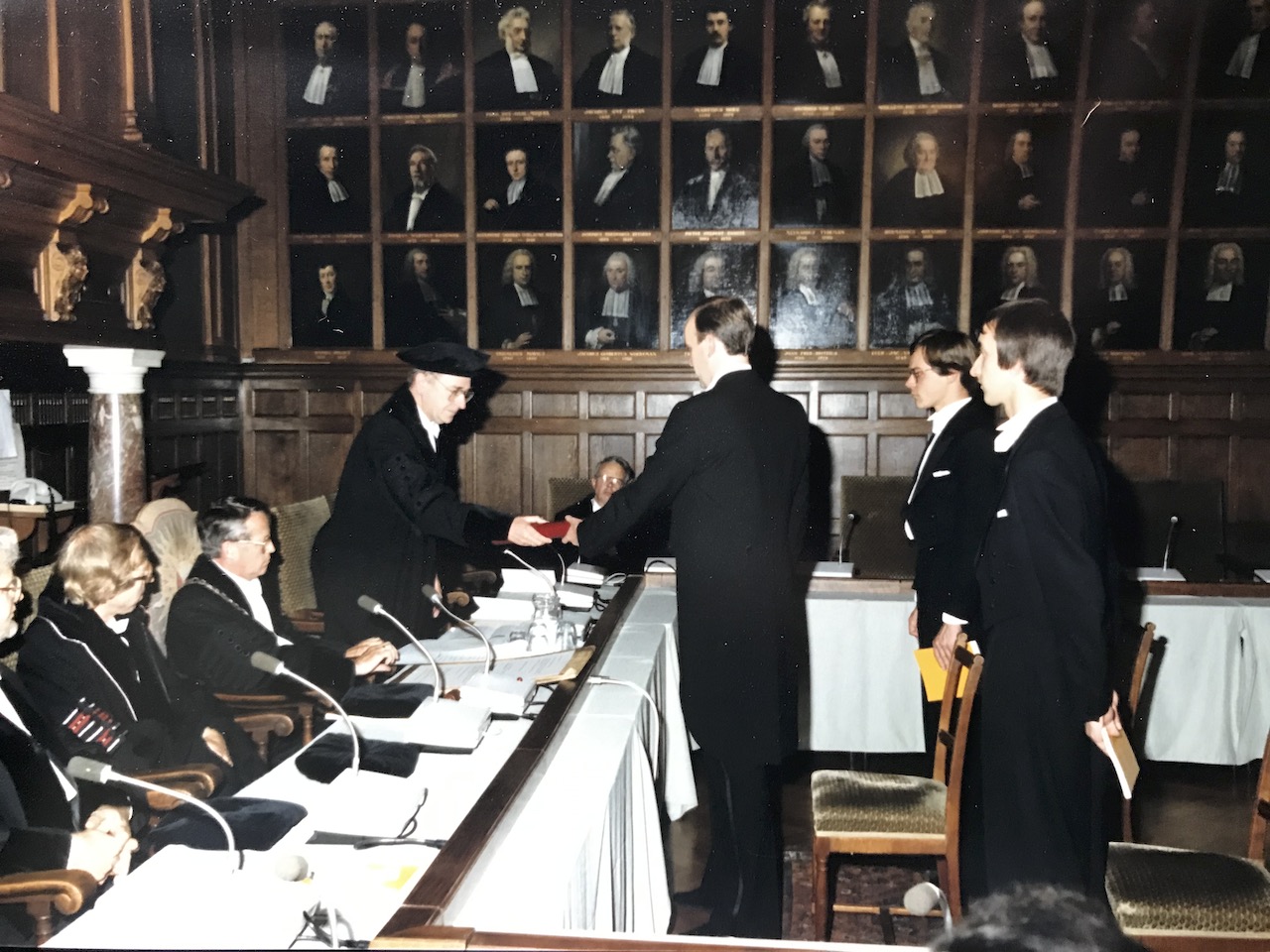}
              }
\caption{Kees Zwaan handing me the diploma for my doctorate in the venerable
Senaatszaal of the Academiegebouw (University Hall) of the University of
Utrecht on 1986/09/22. The title of my thesis, ``Stellar Magnetic Activity:
Complementing Conclusions based on Solar and Stellar Observations,'' formed
the theme of much of my career. Behind me are my paranymphs (ceremonial
assistants): (left) Ren\'e Rutten with whom I worked closely throughout
our Ph.D. phases and (right) Pieter Mulder with whom I go back to freshman
physics. Astronomical research in Utrecht ended in 2012 when the Astronomical
Institute was closed, 370 years after Utrecht became the second city in
the country (after Leiden) with an observatory.}
\label{fig:phd}
\end{figure*}
 In the meantime, Kees Zwaan and Rolf Mewe had reached a decision: they
offered me a Ph.D. position, to be shared between the Space Research Organization
and the Astronomical Institute.

\section{Moving About}
\label{sec:employment}

\textbf{1981\,--\,1986:} \emph{Astronomical Institute Utrecht and Laboratory
of Space Research (LRO).} In those days in the Netherlands, the Ph.D. phase
consisted exclusively of research. There were no classes to take or give,
and no teaching assistantships to take on. I~moved back and forth between
two offices: one at the modern office building of LRO in the outskirts
of Utrecht and one in the Sonnenborgh Observatory\footnote{\url{https://www.sonnenborgh.nl/bezoekersinformatie}.}
in the old city center. At LRO, there were a dozen Ph.D.  students, all
in about the same phase, forming a group of friends who to this day still
occasionally get in touch and who, for years, undertook group outings somewhere
around the country with partners and, later, often their children. At the
Observatory, in line with Dutch tradition, there were the daily mid-morning coffee
breaks and mid-afternoon tea breaks where staff and students
gathered in the small picturesque library for scientific discussions and
small talk.

Toward the end of my Ph.D. appointment (Figure~\ref{fig:phd}), I applied
for, and was offered, a postdoctoral position with Carole Jordan, a solar
and cool-star spectroscopist in Oxford. Had I gone that way, my career
would doubtlessly have taken a turn into spectroscopy, but it was not to
be. Until that year, compulsory military draft would be postponed if one
had a job abroad, but then the rules changed to require a job outside of
Europe. I~was not at all motivated to join the army (considering armies to be
tools of suppression), even less so because those born a year earlier or
a year later than I were excused as the army was transitioning into an
all-professional service. Fortunately, an alternative materialized when
Jeff Linsky (an early leader in cool-star physics and in the study of
asterospheres and the interstellar medium at the Joint Institute for
Laboratory Astrophysics in Boulder, Colorado) happened to visit Utrecht:
my thesis advisor, Kees Zwaan, introduced us, and I was interviewed on
the spot, cornered on a sofa by Jeff and Kees. Three months later, I joined
Jeff's cool-stars group in Boulder.

\textbf{1986\,--\,1988:} \emph{JILA, University of Colorado, and visiting scientist
at Lockheed Palo Alto Research Laboratory (LPARL) and the National Solar
Observatory's (NSO's) Sacramento Peak Observatory.} Boulder was an incredibly
stimulating environment, primarily because Jeff Linsky's group (``the Cool
Star Mafia'') --~at the time including Tom Ayres, Tim Bastian, Jay Bookbinder,
Alex Brown, Bob Dempsey, Jim Neff, Steve Saar, Fred Walter, and Brian Wood~--
provided a very welcoming and broadly knowledgeable work environment. Boulder
is also blessed with an amazing natural setting that I often explored with
long mountain hikes and drives.

Jeff gave me a remarkable amount of freedom in my research into solar and
stellar magnetic activity, and also allowed me to visit other research
groups. After the first year in Boulder, I packed everything I owned into
my car (fortunately, I had been wise enough to invest in a hatchback that
even accommodated my bicycle) and drove Route 50 (much of which is ``the
loneliest road in America'') through mountains and deserts to the San Francisco
Bay Area, which welcomed me with fog rolling in spectacularly over the
Bay and with jammed traffic.

There, I spent three months with Alan Title's group in Palo Alto, working
on supergranular diffusion and the decay of active regions. As much as
my move to Boulder, this was another pivotal moment in my life. Not only
did it set the stage for my later return to Lockheed, I also fell in love
with Iris during this visit. She had arranged to go on a tour of the American
Southwest with my sister, but as Iris had a longer summer holiday, I invited
her to come over early, giving me a traveling companion to explore northern
California. Two years later, we were married and she's been my life companion
ever since.

After that, I spent a month at Sac Peak at the invitation of Rich Radick,
working on relationships between radiative losses from the outer atmospheres
of cool stars. I~loved it there, both because of the people and the setting;
given an apartment in what once were the Visiting Officers Quarters right
at the edge of the escarpment, I had a spectacular view of the wide valley
holding the White Sands Missile Range, the White Sands National Park, and
the San Andres Mountains, with summer lightning storms flashing through
the distant sky almost every night I was there. From there, I moved back
to Boulder to continue in Jeff's cool-stars group.

\begin{figure*}
   \centerline{\includegraphics[width=0.70\textwidth,trim=0mm 40mm 0mm 40mm,clip=]{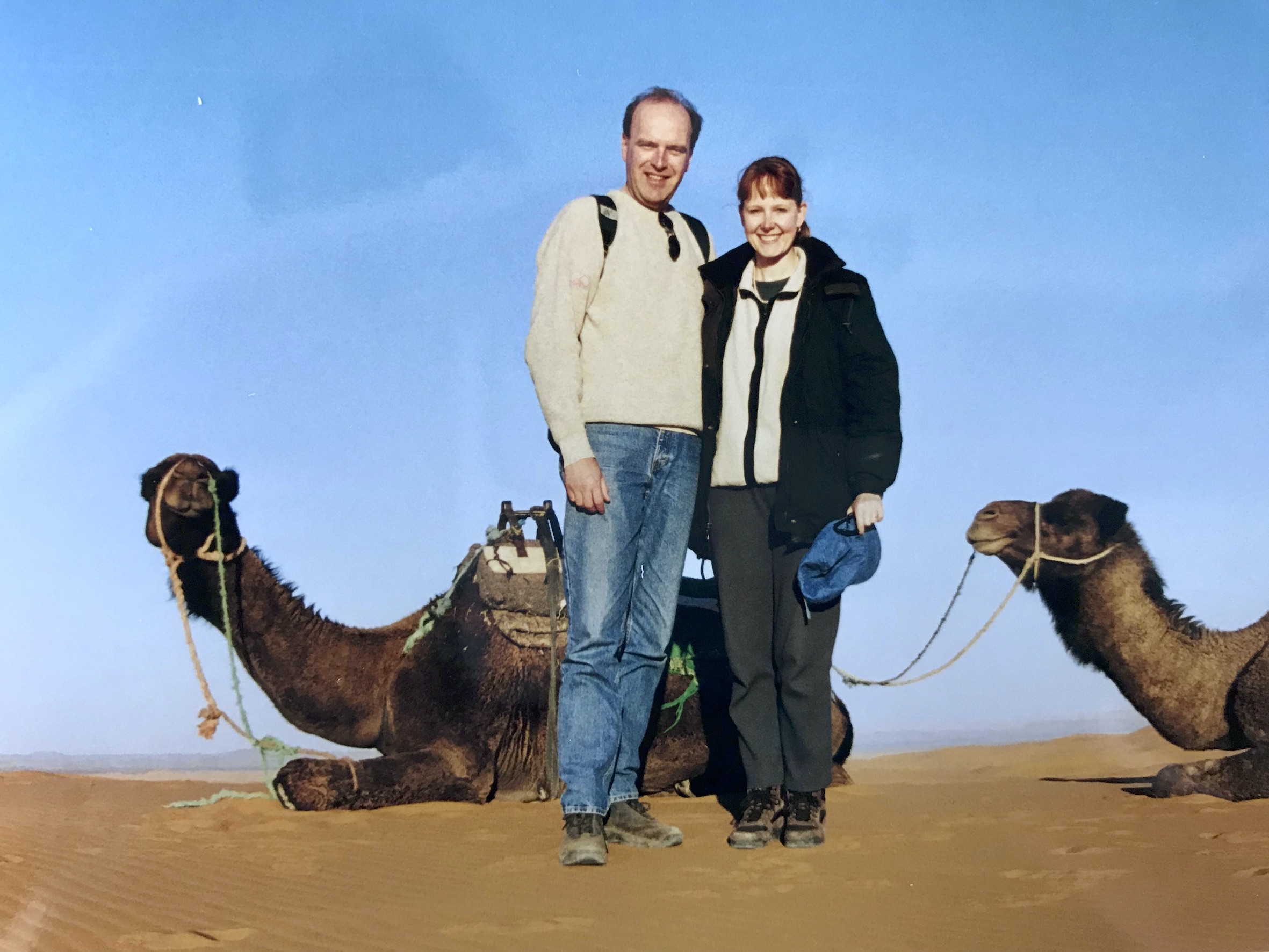}
              }
 \caption{Iris and I both love traveling. Here we are in the Sahara, deep
inside Morocco, somewhere near Mhamid close to the border with Algeria
toward the end of a long tour around the northern edge of the Mediterranean.
We cannot decide on which sights impacted us most in all our travels, but
we did notice a shift over the years in our interest from historical and
cultural sites to the wildest, most remote places on Earth.}
\label{fig:Sahara}
\end{figure*}
\textbf{1989\,--\,1991:} \emph{European Space Agency (ESA), ESTEC, Noordwijk.}
With Iris embarking on her medical training, I looked for a job in the
Netherlands. I~nearly ended up installing flow-monitoring systems along
rivers in China when a postdoc appointment at ESA's European Space Research
and Technology Centre (ESTEC) in Noordwijk came through. It was an opportunity
to learn about helioseismology. I~was tasked to make something of months-long
records of solar oscillations observed in the irradiance signal onboard
the Russian Mars-bound \textup{Phobos} mission. It carried a small instrument
that had a poorly baffled tube looking back at the Sun and therefore recorded
strong internal reflections as the spacecraft's pointing meandered about
the Sun. Working in Noordwijk made me appreciate the beauty of the fields
of flowers along which I commuted. While at ESTEC, military service called
once more, but as the army would have to provide me with a salary matching
my tax-free income at ESA because spouses were not supposed to suffer hardship,
they gave up on me, this time for good.

\textbf{1991\,--\,1994:} \emph{Fellow of the Royal Academy of Arts and Sciences
(KNAW) at the Utrecht Astronomical Institute.} As my appointment at ESTEC
ran out, I applied for a KNAW fellowship, which gave me a free choice in
research topics and where I would choose to work. It wasn't a hard decision:
I went back to my Alma Mater, to Utrecht's Astronomical Institute, to work
with Kees Zwaan, Rolf Mewe, Bert van den Oord, and others, while able to
commute to and from work with Iris. Before Iris started her internships
and I my KNAW fellowship, we took a four-month ``sabbatical'' to literally
travel around the world, taking in some of the beauty that our planet has
to offer. Our love of traveling has never left us (Figure~\ref{fig:Sahara}).

\textbf{1995\,--\,2016:} \emph{Lockheed Martin Advanced Technology Center.}
In the meantime, a third major pivotal event in my career was unfolding. I~had applied for the directorship of the Kiepenheuer Institute in Freiburg,
Germany. Told it would be a good experience to go through such an exercise,
I was rather surprised when I was offered the job, particularly given the
status and experience of the other applicants. It was not to be.
Without disclosing why, someone in the government of the state (``Land'')
of Baden-W\"urttemberg decided to modify the job offer to one that I
could not accept (I would have understood reservations about my lack of
management experience but it appears that not being German played a major
role). Although disappointed at the time, I now think it was the right
outcome: instead of heading to Freiburg, I accepted the standing invitation
to join Alan Title's group in California, a decision I have never regretted.

\begin{figure*}
  \centerline{\includegraphics[width=0.75\textwidth,trim=0mm 10mm 0mm 70mm,clip=]{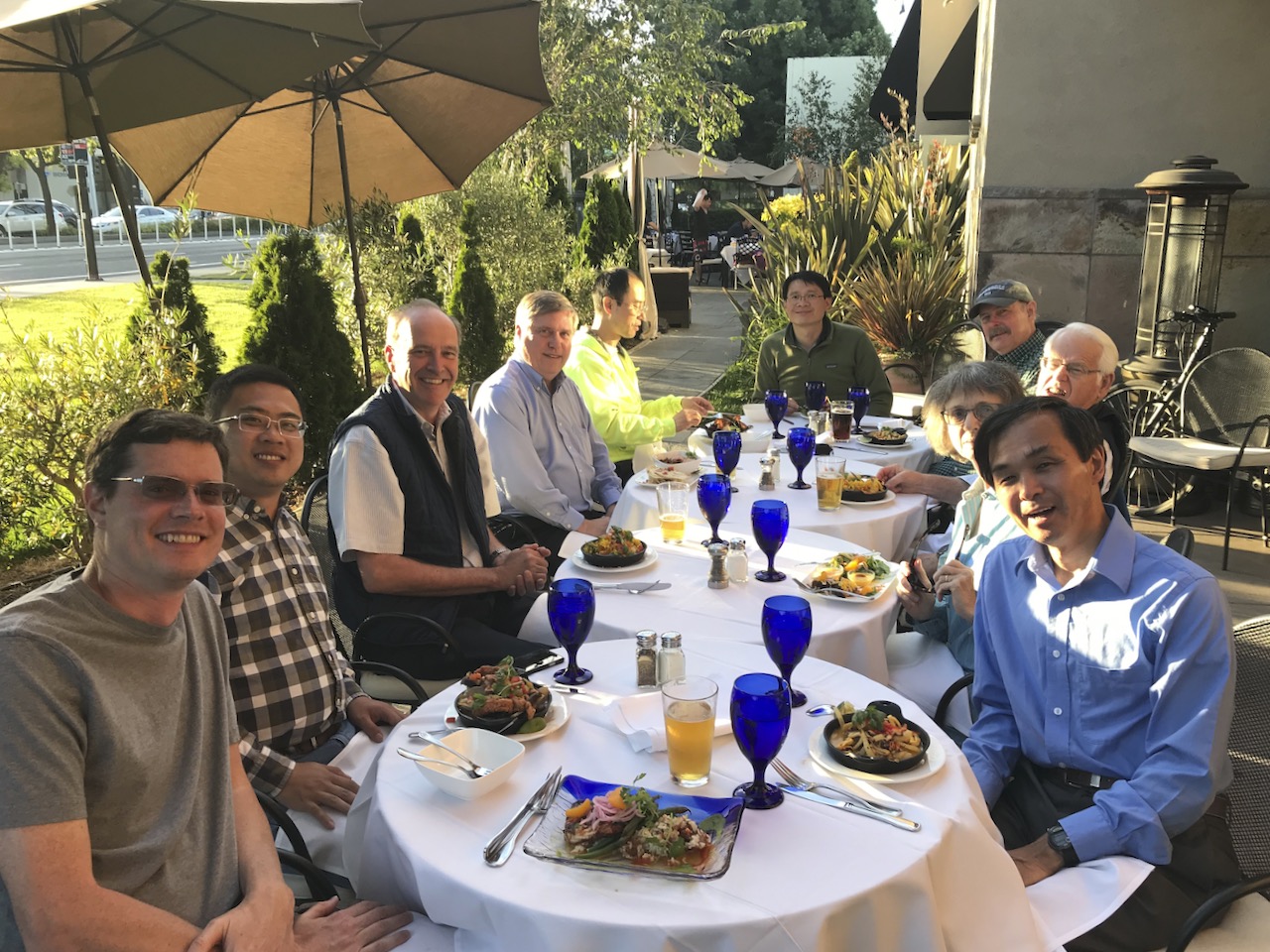}
              }
 \caption{Friends at Lockheed in 2017. Clockwise around the tables: Marc
DeRosa, Meng Jin, me, Neal Hurlburt, Wei Liu, Mark Cheung, Ted Tarbell,
Alan Title, Ruth Peterson, and Nariaki Nitta.}
\label{fig:friends}
\end{figure*}
 There was, however, a challenge to deal with in moving to the USA. As
Iris had just finished her medical training, I was essentially asking
her to redo all her qualifying exams and start over in a different language,
culture, and healthcare system. We made a deal: if she would not find suitable
employment within a year or two, we would move back to the Netherlands.
That agreement was never called upon. Fortunately, she found a job at Stanford
where she ultimately became a full professor in molecular genetics and
clinical pathology. We still look back in amazement at the odds of being
able to combine our two careers while working at most a few miles apart.

I joined Lockheed's Solar and Astrophysics Laboratory (LMSAL) in 1995
as the Transition Region and Coronal Explorer (\textup{TRACE}) was entering
its construction phase and when the science planning was ramping up. It
was a steep learning curve for me, transitioning from a user of spacecraft
observations to being involved in planning and --~after the 1998/04/01
launch~-- executing observations. In that period, I became the science
lead. In 2005, when Alan Title started as Principal Investigator of the
Atmospheric Imaging Assembly (AIA) onboard the \textup{Solar Dynamics Observatory}
(\textup{SDO}), I took over his role as PI of \textup{TRACE}. I~learned a lot
from inspecting \textup{TRACE}'s crisp coronal and transition-region imagery
almost every workday for twelve years. Also, I learned a lot about working
with NASA and the international scientific community.

\begin{figure*}
   \centerline{\includegraphics[width=0.75\textwidth,trim=0mm 60mm 0mm 30mm,clip=]{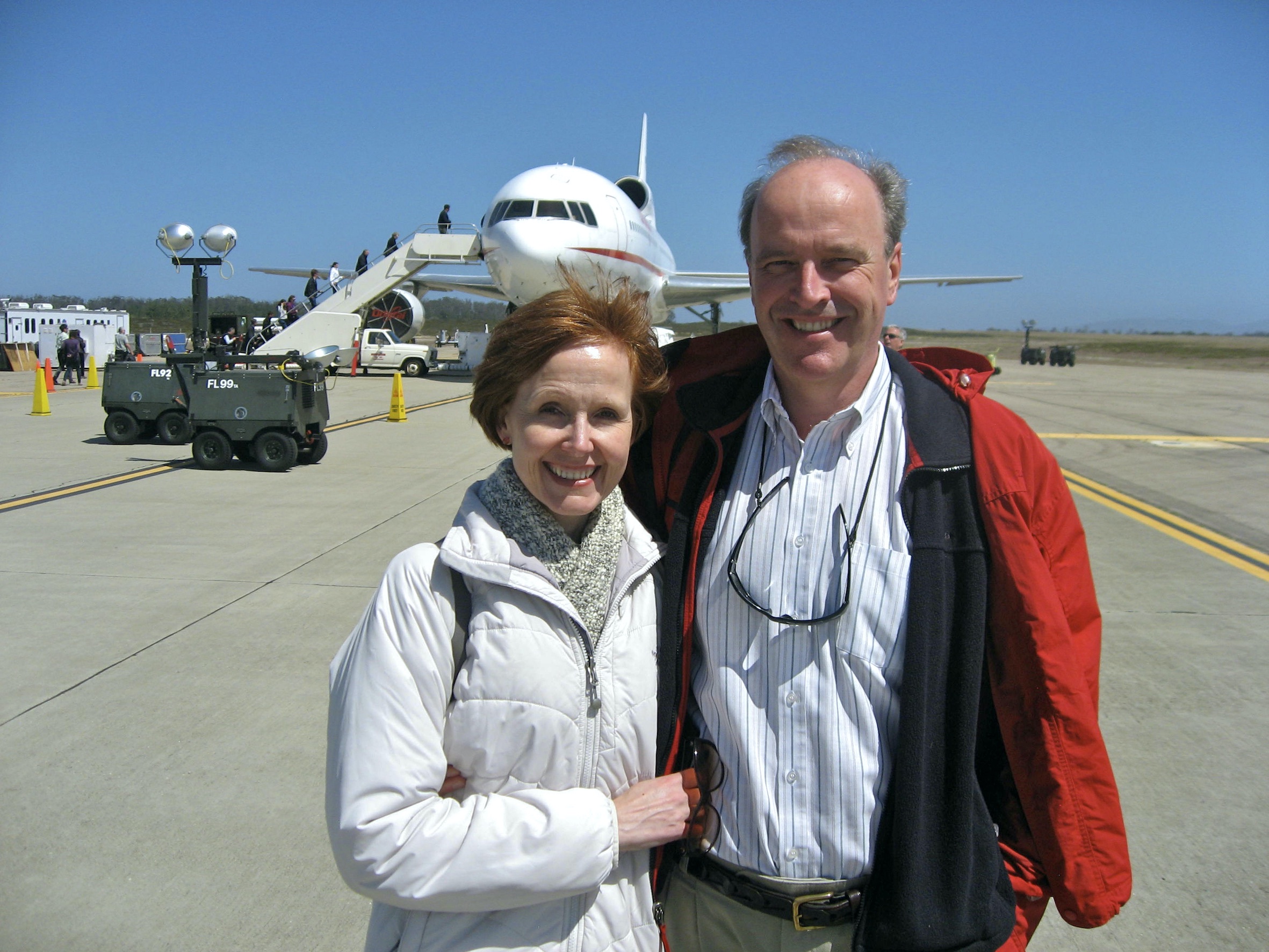}
              }
 \caption{Iris and I visited Vandenberg Air Force Base (now called a Space
Force Base) on 2013/06/22, a week before the launch of the \textup{Interface
Region Imaging Spectrograph} (\textup{IRIS}). Behind us is the L-1011 TriStar
aircraft with \textup{IRIS} already attached to the Pegasus rocket mounted
underneath the plane. Whereas I was very involved in the development and
proposal phases of \textup{IRIS}, I, unfortunately, lacked the time to engage
in its research.}
\label{fig:IRIS}
\end{figure*}

More of those lessons followed when LMSAL (Figure~\ref{fig:friends}) was
awarded the contract to build \textup{SDO}'s AIA. \textup{SDO} was launched successfully
on 2010/02/11. Once its torrent of telemetry came down, it gave us an essentially
continuous, high-cadence, multitemperature view of the entire Earth-facing
part of the solar outer atmosphere, revealing its dynamics in unprecedented
detail. We retired \textup{TRACE} soon after AIA was demonstrated to be fully
functional. Just short of \textup{SDO} completing its second year on orbit,
I transitioned from AIA's science lead to PI as Alan started work on the
\textup{Interface Region Imaging Spectrograph} (\textup{IRIS}; Figure~\ref{fig:IRIS}),
adding another six years of near-daily inspection of the solar corona before
I retired in 2016.

At this point, I will end the chronological summary of my life, instead
following the development of selected scientific themes.

\section{Stars Like the Sun and the Sun-as-a-Star}
\label{sec:sunstar}

A substantial part of my career focused on what is generally referred to
as ``the solar--stellar connection.'' A full-text search with the NASA
Astrophysics Data System (ADS) (what a wonderful tool to have!) suggests
first use of that term by Andrea Dupree (a stellar astrophysicist with
focus on cool-star science) in the preface to the proceedings of the very
first workshop
\citep [held in Cambridge, Mass.;][]{1980SAOSR.389D...7D} in an ongoing
series entitled ``Cool Stars, Stellar Systems, and the Sun.'' That first
meeting stimulated the research in the field, further pushed along by a
meeting on the other side of the Atlantic the following year, a NATO Advanced
Study Institute organized by Roger Bonnet (who would later become ESA's
director of science and of ISSI, among other things) and Andrea Dupree
\citep{1981ASIC...68.....B} that is widely referred to as ``the Bonas meeting.''
In the introduction to its proceedings, the editors wrote ``In the past,
stellar and solar astrophysics have more or less followed their own independent
tracks. However, with the rapid development of modern techniques, in particular
artificial satellites like the \textup{International Ultraviolet Explorer}
[\textup{IUE}] and the \textup{Einstein} Observatory, which provide a new wealth
of data, it appears that chromospheres, coronae, magnetic fields, mass
loss and stellar winds, etc. {\ldots}, are found not only in the Sun but
occur also in other stars. Frequently these other stars represent quite
different conditions of gravity, luminosity, and other parameters from
those occurring in the Sun. The Sun is no longer an isolated astrophysical
object but serves the role of representing the basic element of comparison
to a large class of objects.''

I did not attend either of these two meetings but their proceedings formed
most of my Ph.D.'s startup material. The above statement by Bonnet and
Dupree could have served as the rationale for my thesis work, which started
for me with a 3-month pilot project (or Ph.D. probationary period, depending on perspective) in 1981 that launched me
on a decades-long study of the relationships between radiative losses from
different thermal regions in the atmospheres of Sun-like stars.

\subsection{Power Laws and Basal Fluxes}
\label{sec5.1}

Thus, even before formally graduating, I embarked on my first trip to the
USA for an extended visit to the Center for Astrophysics (CfA, a collaboration
between the Smithsonian Astrophysical Observatory and the Harvard College
Observatory, with boundaries running through the building complex that
never became clear to me) in Cambridge, Mass. Rolf Mewe accompanied me
for that first visit for a week to introduce me to people, places, data,
and codes. Kees Zwaan and Rolf Mewe already had a working relationship
in place with the \textup{Einstein} project \citep{1981ApJ...245..163V} and
thus gave me a running start.

Rolf had found a boarding house, run by a former secretary of the CfA.
The house was almost a real Victorian: all-wood construction (a contrast
to the predominantly brick or concrete buildings in the Netherlands), with
creaking floorboards, insufficient ventilation on humid summer days and
--~during later winter visits~-- steam running through the small radiators
in an attempt to fight the loss of heat through the single-pane windows.
We would go out for breakfast (another surprise for me: you couldn't go
out for breakfast in the Netherlands at the time, except perhaps in a hotel)
at a classic American diner and then we would walk up to the CfA, perched
on its hill.
\par
We would sit down at the monitors, and fit spectra from the spectral code
that Rolf had been developing \citep{1972SoPh...22..459M} to \textup{Einstein}'s
observations of select cool stars. There was little information to extract
from the signals from the imaging proportional counter: generally a single
temperature for the stellar corona combined with a column density of interstellar
absorbing hydrogen yielded a perfectly acceptable fit, although it was
tempting to try two-temperature approximations. The main findings were
the marked distinction in dominant coronal temperatures between main-sequence
stars (around 3 MK) and evolved stars (in the 10\,--\,20 MK range), and the
initial establishment of the relationship between chromospheric Ca\,{\textsc{II}}
H$+$K signals in the visual as recorded at Mt. Wilson and the coronal soft
X-ray brightness measured by \textup{Einstein} \citep{ref45}, published shortly
after \cite{ref180} had explored such relationships for UV and coronal
diagnostics.

As my Ph.D. phase began, I joined Kees Zwaan's research group, then with
Frans Middelkoop
\citep[who had made several observing runs at Mt. Wilson, working with Olin
Wilson on his project to measure Ca\,{\textsc{II}} H and K line strengths
in cool stars that had its roots in][]{ref84} and Barto Oranje (who focused
on \textup{IUE} spectra of cool stars and on Sun-as-a-star observations with
the Utrecht solar spectrograph), but soon after also with Ren\'e Rutten
with whom I worked most closely. Initially, I spent my time mostly at LRO
across town, apart from weekly meetings in Kees Zwaan's spacious office
at the Observatory. Over time, though, I worked increasingly in the office
over the entrance to the Observatory that I shared with Barto, Frans, and
Ren\'e. It offered wonderful views of the majestic old trees in the adjacent
park where I often spent lunchtime breaks walking along the Singel --~a
wide defensive moat running at the base of the Observatory~-- unless I
opted for the old center of Utrecht.

Those were electrifying days: the \textup{IUE} and \textup{Einstein}/HEAO-2 spacecraft
had been launched in 1978 to observe, among other things, cool stars like
the Sun, followed in 1980 by the \textup{Solar Maximum Mission} (\textup{SMM})
on the solar side. Discoveries of coronal mass ejections
\citep[CMEs, first observed
with \textup{OSO-7} and then in more detail from the \textup{Skylab}
space station;][]{1973spre.conf..713T,1974ApJ...187L..85M}, of the relationship
between chromospheric Ca\,\textsc{II} emission, stellar age, and rotation
(\citealp{ref133}, see \citealp{2018A&A...619A..73L} for a recent
confirmation of his results for a sample of solar twins), and the
principles of magnetic braking of stellar rotation
\citep{1968MNRAS.138..359M} were all hardly a decade old and the research
into these phenomena was in full swing. It did not take me long to realize
that I lived in an exciting time and was embedded in a wonderfully pivotal
research group, being guided by the experienced observational solar physicist
Kees Zwaan and the plasma spectroscopist Rolf Mewe, working alongside Ph.D. students
using ground- and space-based observatories in a rapidly growing research
field.

Establishing and understanding relationships between radiative losses from
different physical domains in the outer atmospheres of cool stars was my
initial focus area (working in parallel to, primarily, Jeff Linsky's group
in Boulder --~e.g., \cite{ref180}~-- whose \textup{IUE} observations were
essential to our early research, complemented by our own data). As Ren\'e
Rutten and I were assembling data from Mt. Wilson, \textup{IUE}, and \textup{Einstein},
I noticed something in the diagrams, namely a dependence on the stellar
spectral color similar to what we had seen when relating coronal soft X-rays
and chromospheric Ca\,{\textsc{II}} H$+$K. That color dependence disappeared when
I subtracted what I named the ``basal flux,'' an emission level running
right below the lowest detections of the stellar signals in flux-color
diagrams.

For the Ca\,{\textsc{II}} H$+$K signal the presence of a minimal level was to
be expected because the Mt. Wilson spectrograph also transmitted a contribution
from the line wings. However, for the UV, with little or no continuum radiation
underlying the emission lines for many cool stars, it did come as a surprise.

My thesis supervisor and my fellow Ph.D. students were skeptical at first.
They had not seen a need to introduce such a basal level in the \textup{IUE}
UV data themselves, and neither did the Boulder group, which was also producing
flux--flux diagrams at the time. However, as more and more data came in,
the evidence convinced us. Once the basal level was removed, flux--flux
relationships held up for giants and dwarf stars alike, regardless of surface
gravity or temperature
\citep[except, as we noticed later on, for
the very cool M-type dwarfs that are deficient in chromospheric
emission,][]{ref8}. This saw me embark on a 14-year research thread that
eventually culminated in a review \citep{ref312} concluding that ``There
is substantial quantitative observational and theoretical evidence that
the basal emission from stellar outer atmospheres is caused by the dissipation
of acoustic waves generated by turbulent convection'' although ``effects
of intrinsically weak fields on the basal mechanism cannot be entirely
ruled out.'' A decade later, noting an error in my calibration of solar
to stellar data for the high-chromospheric C\,{\textsc{II}} signal, using new
observations made with \textup{SoHO}'s SUMER instrument, and supported by
state-of-the-art radiative hydrodynamic modeling,
\cite{judge+etal2003} stated that they ``believe (but cannot prove) that
[their] analysis points to [internetwork] magnetic fields as the dominant
origin of the basal component seen in C\,{\textsc{II}} lines.'' Even as of today,
more observations and modeling are required to sort this out: is all of
the basal flux related to weak turbulent fields, or is there a shift from
acoustic to magnetic going up in temperature and height, or is it all dominated
by the dissipation of mechanical energy?

Reaching the conclusion that basal fluxes are a reality (regardless of
their physical origin) would take me through several projects as well as
to interesting collaborations and locations in my postdoc years. Among
these was, for example, a study of the transition-region C\,{\textsc{IV}} emission
as observed by \textup{IUE} for cool stars, for the Sun-as-a-star by the \textup{Solar
Mesosphere Observer} (\textup{SME}), and with spatial resolution by the \textup{SMM}-UVSP
(with help from Dick Shine at Lockheed), in comparison with magnetogram
data (recovered from the NSO archive with the help of Stuart Ferguson and
Jack Harvey). Then, there was a project for which Rich Radick invited me
to Kitt Peak to work with him and Andrea Dobson on relating stellar observations
of Ca\,{\textsc{II}} H$+$K, Mg\,{\textsc{II}} h\&k, and stellar rotation rates. A~few years later, Rich invited me once more for a study, again with Andrea,
on establishing flux--flux relationships between stellar Ca\,{\textsc{II}} H$+$K,
Mg\,{\textsc{II}} h\&k, and soft X-ray fluxes using observations with the
smallest available time differences between these various observations,
which, unfortunately, still ranged from a day or two to in excess of a
month even for the best cases --~clearly not ideal.

Allow me to make a critical comment here: it has always baffled me, and
it continues to baffle me, that the coordination between various space-
and ground-based observatories is so exceedingly difficult to achieve.
Doing so between different countries is particularly challenging, but within
a country with the resources managed by only a few organizations --~such
as within the US primarily by NASA and NSF~-- it should be readily feasible. I~mean, we have computers to sort out schedules, and if airlines can do
so most of the time for most of its passengers, surely astronomers should
be able to pull it off for a handful of observatories. However, as long
as funding streams keep requiring multiple proposals for multiple observatories
and time-allocation committees are not coordinated, we shall continue to
waste promising opportunities and a lot of researchers' time.

In this context, there is, unfortunately, also the dominant role of personal
interests to contend with: I just could not understand why, for example,
even the instruments on a single spacecraft such as \textup{SoHO} often did
not focus on the same region on the Sun. When I was planner for \textup{TRACE},
or in my role as its science lead and later PI, I would often opt to follow
the lead of one of \textup{SoHO}'s observers, and then frequently stick with
a target for days on end. I~think that paid off as witnessed by the many
multi-instrument papers written involving \textup{TRACE} observations. Moreover,
with this strategy, we also managed to capture the majority of the most
energetic events on the Sun.

I can only hope things will improve because there is such a promise of
discovery in joint observations. I~think it was Jeff Linsky who came up
with the word ``panchromatic'' as he pointed out the value of such observations
decades ago. Also, I appreciated it first hand when observers combined
data from six international spacecraft, covering the wavelength range from
the optical into gamma rays, to study the initiation and evolution of the
powerful X17 flare SOL2003-10-28T09:51 \citep{schrijver+etal2005c}.

\begin{figure*}
  \centerline{\includegraphics[width=0.75\textwidth,trim=20mm 130mm 20mm 25mm,clip=]{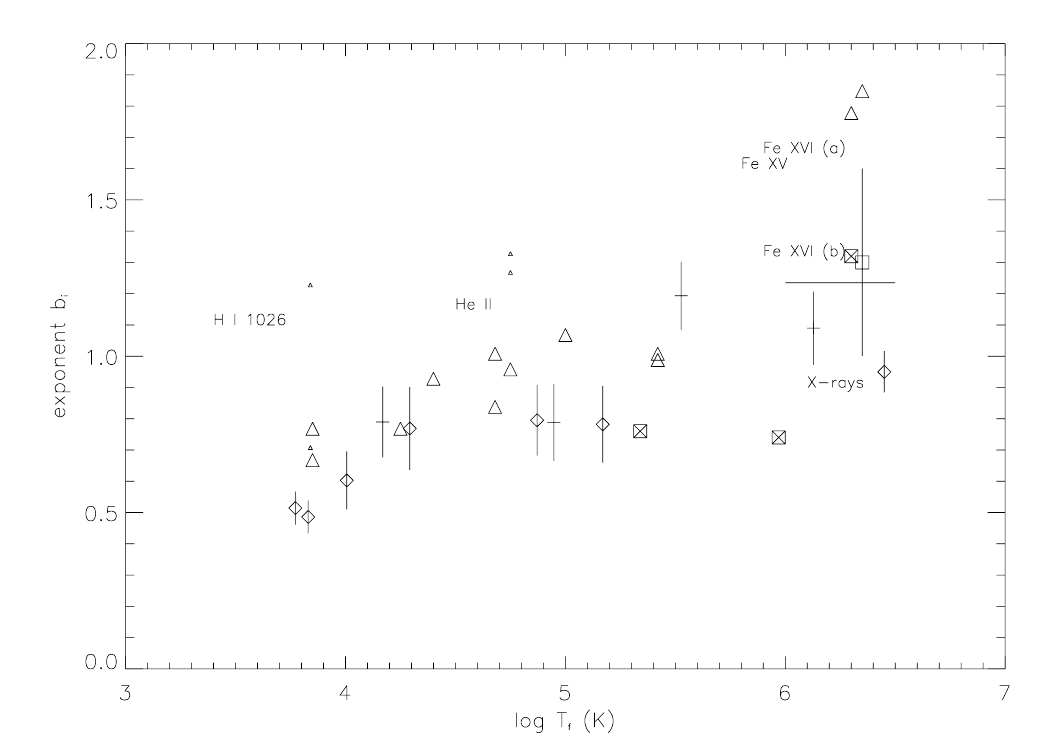}
              }
 \caption{Radiative losses $\Delta F_{\mathrm{i}}$ in excess of any basal emissions
from the outer atmospheres of cool stars are related to each other and
to the underlying magnetic flux densities
$\langle | \phi | \rangle $ through power laws independent of stellar color
or luminosity:
$\Delta F_{\mathrm{i}} \propto \langle | \phi | \rangle ^{b_{\mathrm{i}}}$. This
diagram is a compilation of power-law indices $b_{\mathrm{i}}$ for radiative
diagnostics with formation temperatures that range from chromosphere to
corona, revealing the tendency towards increasing steepness with temperature.
(From \citealp{schrijver+siscoe2010b}.) This diagram is similar to Figure~7
in \cite{2022ApJS..262...46T}, although the slopes for the chromospheric
diagnostics in that study are higher, which may have to do with their definition
of the ``basal'' level to be subtracted $^{\ref{foot:basal}}$.}
\label{fig:ffexp}
\end{figure*}
%

\subsection{Sun-as-a-Star Construction Kit}
\label{sec5.2}

Let's go back to my Ph.D. years. After correction for the basal emission, radiative losses from chromospheres
through coronae of all sorts of cool stars defined power laws (Figure~\ref{fig:ffexp}).\footnote{Differing
power-law indices have been reported for the relationships between a given
radiative signal and magnetic-flux densities for a variety of spectral
lines (although authors generally agree that the power-law index tends
to increase with formation temperature from the chromosphere upward). Among
the causes for these differences are the variety of definitions of what
the authors consider the basal level (solar network interior or average
network or lowest magnetic signals or minima from stellar samples, with
uncertainties in the conversion between stellar flux densities and the
solar intensities) and the sensitivity of the results to bandwidths (with
different scalings reported within profiles of optically thick diagnostics);
see, e.g., \cite{2018A&A...619A...5B,2022ApJS..262...46T}. In my work,
the basal emission is that emission level that removes any dependence on
fundamental stellar parameters from flux--flux relationships.
\label{foot:basal}%
} Of course, I was intrigued by the nonlinearity of these flux--flux relationships
that held over some four orders of magnitude in soft X-ray flux density,
regardless of which type of star we looked at within the late-F, G, K,
and early-M type dwarfs and giants. The fact that the Sun appeared to move
largely along these relationships going from solar minimum to maximum triggered
the concept of a digital ``construction kit'' \citep{ref232}. Using the
solar cycle as a guide, I placed different numbers of active regions on
a simulated solar surface, included mixed-polarity quiet Sun, and thus
studied the behavior of the Sun as a whole. That required, of course, knowledge
of the constituent parts: the quiet Sun, the variety of active regions,
and their radiative losses.

It started easily enough. I~looked through the listings of the Solar Geophysical
Data to determine the frequency of active regions existing on the solar
surface at any instant in time as a function of their Ca\,{\textsc{II}} K plage
area \citep [following][who differentiated cycle minimum and maximum phases using Mt. Wilson magnetograms]{1984SoPh...91...75T}. I~remember Kees Zwaan's surprise at the smoothly decreasing curve with
increasing region size: until presented with this data, he had the
impression that there were distinct populations of small and large active
regions. Moreover, the frequency distribution appeared to change through
the solar cycle only by a multiplicative factor.

\begin{figure*}
  \centerline{\includegraphics[width=0.59\textwidth,trim=20mm 120mm 40mm 20mm,clip=]{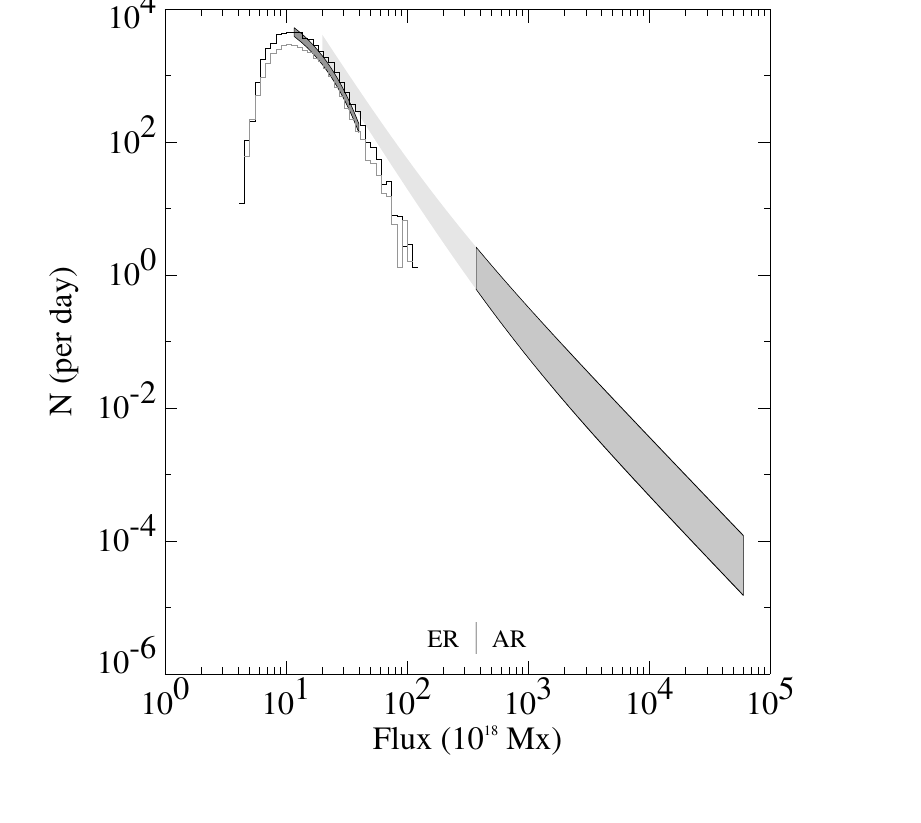}
              \includegraphics[width=0.41\textwidth,trim=20mm 120mm 110mm 50mm,clip=]{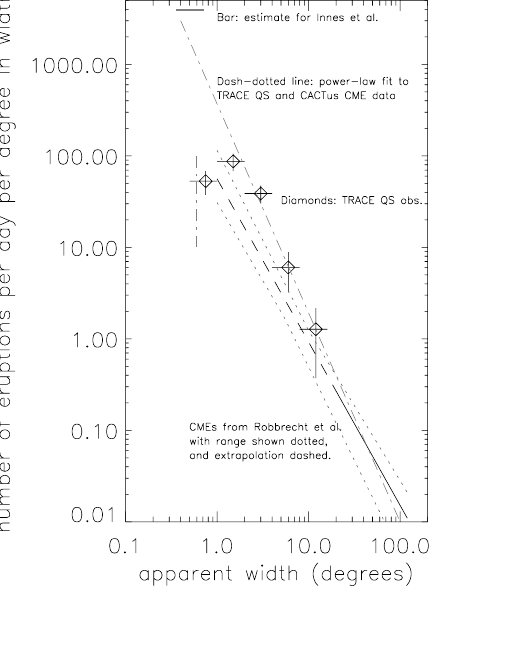}}
 \caption{The unsigned flux distributions for bipolar regions (\emph{left})
and the angular widths of eruptions (\emph{right}) appear to smoothly connect
from the small ephemeral-region scales to active regions and their CMEs.
\emph{Left}: Frequency distribution of bipoles emerging on the Sun per day,
per flux interval of $10^{18}$ Mx. The number of active regions varies
by about a factor of 8 through a typical cycle (indicated by the shaded
area on the right) while the variation of ephemeral regions is much smaller
(and possibly in antiphase). The turnover below $10^{18}$ Mx likely reflects
the detection threshold of \textup{SoHO}/MDI. The range between ephemeral
regions and active regions still awaits observational confirmation.
(From \citealp{hagenaar+etal2003}.) \emph{Right}: Frequency of observed eruptions
for \textup{TRACE}, \textup{SoHO}/LASCO, and \textup{STEREO}/EUVI versus angular
width. (From \citealp{schrijver2010}.)}
\label{fig:powerlaws}
\end{figure*}
Karen Harvey (also a graduate student of Kees Zwaan, defending her thesis
in 1993) would later take a somewhat different approach to come to a comparable,
although distinct, conclusion \citep{1993SoPh..148...85H}. Instead of determining
how many regions of what size existed on the solar surface at a given time,
she painstakingly reviewed daily Kitt Peak magnetograms to determine how
many newly emerging regions ultimately grew to what size at full development
(Figure~\ref{fig:powerlaws}), thereby removing lifetime effects from my distribution
to establish a true source function (I return to that in Section~\ref{sec:dispersal}).
Here, too, a fixed (in fact, power-law) shape of the distribution emerged,
modulated over time through the solar cycle by a multiplicative factor
(later extended by Mandy Hagenaar to the ephemeral-region range, see
Figure~\ref{fig:powerlaws}). An underappreciated (and remarkably little-known)
property of active regions is that about half of them emerge inside existing
regions: bipolar regions are strongly nested (originally noted by
\cite{cassini1729}, studied in detail by \cite{1993SoPh..148...85H}, and
later by, e.g., \cite{2002SoPh..208...17P}, and inferred for Sun-like cool
stars by \cite{2020ApJ...901L..12I}). It seems that this is a key fact
in how flux escapes from the dynamo source region onto the surface, but
it receives little attention. Comparison of observed rotational modulation
to simulated signals suggests that active-region nesting may be even more
pronounced in more active stars \citep{2023AnA...672A.138N}.

The next step toward the ``construction kit'' was to establish the brightness
of active regions in a variety of spectral lines and bands. I~returned
to the CfA for that, working with Charles Maxson and Bob Noyes. There,
I toiled through the records of the S-055 ultraviolet spectrograph and
the S-054 imaging X-ray telescope on \textup{Skylab}'s Apollo Telescope Mount.
S-054 images were recorded on 70~mm film that was brought down to Earth
by the astronauts, developed, and --~at least for the CfA~-- printed onto
transparencies of some 25~cm in size (as far as I recall), stored in drawer
after drawer in a small room off CfA's hallways. I~plowed through these
to identify different-sized regions at different positions on the disk
with sufficient separation between them to ensure they were effectively
isolated. Then, I worked through the inches-thick stacks of computer output
that held the S-055 log files to determine which of these regions were
scanned in their entirety by that instrument in a set of chromospheric
and coronal lines. The final step in obtaining the data involved finding
a working 7-track tape reader to extract the data from the many rows of
tapes in the archive down the hall.

Collecting that data for two dozen active and quiet regions took two month-long
visits to Cambridge, Mass., followed by a two-week visit to NSO in Tucson
to recover magnetograms for a subset of these for which such data were
available. What a contrast with the modern day in which observing logs
and archives, and even event listings, are available online, often with
search and analysis software provided, for example through SolarSoft\footnote{\url{https://www.lmsal.com/solarsoft/}.}
\citep{1998SoPh..182..497F}. Despite that, population studies
of active-region properties, flares, or coronal mass ejections that combine
data from multiple instruments remain rare. That is disappointing given
the efforts put into enabling such studies by, among others, the science
teams of \textup{SoHO} and \textup{SDO}.

During my stays at CfA, I didn't take nearly enough time off to explore
Boston and its surroundings, but I did append a trek across the USA to
each of these visits, traveling in a small van with 11 like-minded young
people. The first trip took me from New York to Los Angeles, and the second
from Seattle to Miami (with a short stop in Boulder, CO, where I had no
inkling that I would be working there just a few years later). It was during
these trips that I fell in love with the American West. I~was thrilled
by the wide open spaces, by the colorful geology often laid bare to the
eye in desert areas, and by the abundant wildlife (in the small, densely
populated country of the Netherlands, wildlife was largely limited to birds;
I think I saw one squirrel in all the years that I lived there, and only
a few wild boars and deer in an enclosed park, and that only from afar).

Combination of the collected data into the rather crude first ``construction
kit'' (enabling the last chapter of my Ph.D. thesis) showed how the emission
from quiet and active regions combined into Sun-as-a-star signals that
fell roughly, but not quite, along the stellar flux--flux relationships.
Two other conclusions started to take shape: (1) that the main determinant
of the brightness of active regions appeared to be their area, and (2)
that in order for the Sun to fully follow the stellar flux--flux relationships,
the quiet Sun also needed to change with activity level. The first of these
conclusions led to the discovery of the ``plage state'' and Alan Title's
first invitation to Lockheed in the San Francisco Bay area (Section~\ref{sec:dispersal}).

The second of these conclusions led to a month-long visit to the National
Solar Observatory in Tucson. Karen Harvey, a marvelously warm-hearted person,
helped me work through almost a full solar cycle of synoptic magnetic maps
from which we derived distribution functions, ${h}_{t}(|\phi |)$, of magnetic
flux densities across the solar surface over time. We found that the properties
of these time-dependent distribution functions were such that they transformed
power-law relationships between local radiative and magnetic-flux densities,
$f(|\phi |)$, into essentially the same relationships between disk-integrated
signals, $F(\langle |\phi |\rangle )_{t}$, as function of surface-averaged
magnetic flux density, $\langle |\phi |\rangle _{t}$ (here ignoring radiative-transfer
effects):
%
\begin{equation}
\label{eq:fftrans}
F(\langle |\phi | \rangle _{t}) \approx \int {h}_{t}(|\phi |) f(|
\phi |) {\mathrm{d}}|\phi |/\int {h}_t(|\phi|) {\rm d}|\phi|.
\end{equation}
With that, we understood how it could be that the solar spatially resolved
observations and the stellar observations could lead to the same power
laws: the origin lay in the continuous, smoothly changing distribution
of the surface magnetic field, from the quietest areas into the population
of active regions, through the solar cycle. That, of course, led to the
next question: how did that particular type of distribution function come
about? This eventually led to the surface flux-transport model that formed
the basis for the second-generation construction kit, as described in Section~\ref{sec:dispersal}.

\subsection{Linking Radiative Losses to the Magnetic Field}
\label{sec5.3}

The question of how the atmospheric radiative losses connected to the magnetic
field played throughout this research into the nature of flux--flux relationships.
An early answer had been provided by \cite{ref55} who established a linear
relationship between the Ca\,{\textsc{II}} K line brightness and the underlying
magnetic-flux density for a quiet-Sun area. However, that was inconsistent
with what Steve Saar and I found based on stellar data
\citep{1987LNP...291...38S}. Steve, under Jeff Linsky's mentorship, had
developed a method to estimate surface-averaged magnetic-flux densities
on cool stars. When we compared these to coronal soft X-ray data, including
solar measurements, an essentially linear relationship emerged. Through
the relationship between coronal and chromospheric losses, this implied
an approximately square-root relationship between magnetic and chromospheric
flux densities (Figure~\ref{fig:ffexp}), which should hold both for stars
as a whole and for the spatially resolved Sun. To resolve this, I submitted
a proposal to NSO to extend the relationship established by
\cite{ref55} to enhanced network and active regions.

I was awarded two week-long observing runs on the McMath solar telescope
on Kitt Peak. When I arrived in Tucson for the first observing run, I was
given an NSO car and told to drive into the desert on my way to the telescope
along a deserted road lined with saguaros, ocotillos, and sagebrush, encountering
the occasional tumbleweed. I~had never driven an automatic car, in fact
had never driven a car in the USA, and had only just arrived from the Netherlands
with no opportunity to adjust to the time difference, so it was quite the
adventure. That continued once I reached the mountain and was shown how
to use the enormous telescope (guided through the setup by the very patient
and helpful Bill Livingston, Jack Harvey, and Buzz Graves). During the
observing run before mine, something in the telescope building had tripped
the circuit breakers repeatedly, and the observer had simply reset the
breakers each time. This caused the power cable going into the building
to melt. Repair would take time, so NSO ordered a large mobile generator
to power the building. Unfortunately, its running speed varied irregularly,
sometimes causing the telescope's electronics to complain, if not the circuit
breakers to trip. That, the clouds, the rains of the monsoon season, and
the refusal of the Sun to provide an active region when I could have observed
it, made for a failed observing run.

The second run, half a year later, did succeed, with beautiful weather,
a cooperating Sun, and a fully functioning telescope. The data were exactly
what Jacqueline Cot\'e, Kees Zwaan, Steve Saar, and I needed to extend
the relationship by \cite{ref55}. As we had anticipated, their limited
range in flux densities had led them to make a linear fit, whereas a power-law
fit with an index of $\approx 0.6$ lined up the quietest regions to the
active-region core.

\begin{figure}
   \centerline{\includegraphics[width=0.9\textwidth,trim=0mm 30mm 0mm 20mm,clip=]{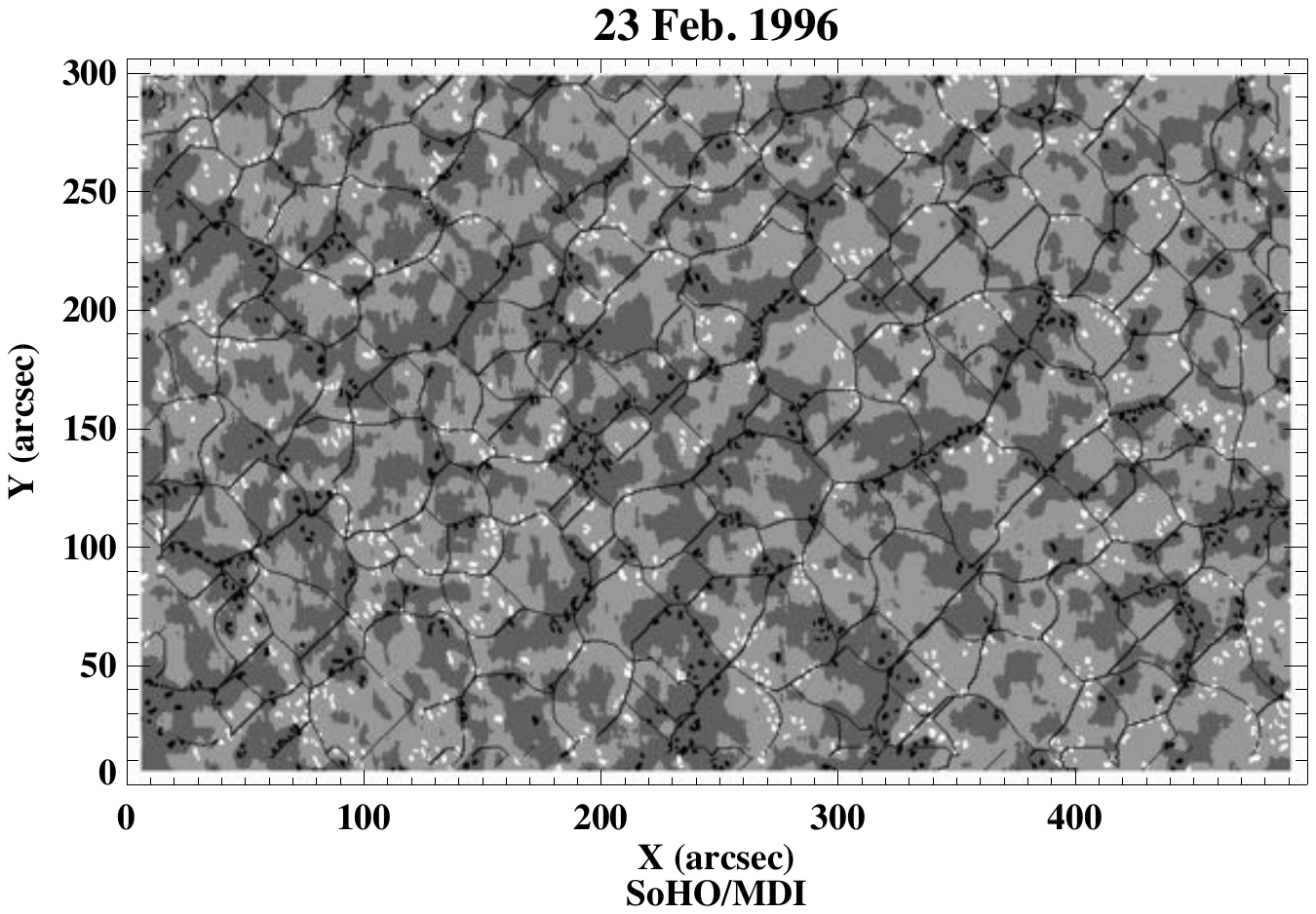}
              }
 \caption{The magnetic carpet, i.e., the mixed-polarity quiet Sun, results
from the continual injection of ephemeral regions, their dispersal in the
supergranular network, and the cancelation of flux in opposite-polarity
encounters. The black and white patches show the flux concentrations for
the two polarities in a \textup{SoHO}/MDI magnetogram, while the light and
dark gray areas show a map of the two polarities after a 6-arcsec smoothing
was applied. The black lines show the cellular network formed by converging
flows, determined from local correlation tracking; this supergranular network
corresponds closely to the chromospheric and magnetic networks.
(From \citealp{schrijver+etal1996e}.)}
\label{fig:carpet}
\end{figure}
%

\subsection{The Magnetic Carpet}
\label{sec5.4}

The analysis of how flux elements are dispersed across the solar surface
led to excursions into percolation theory and tesselation patterns and
from there to a zero-dimensional model of the distribution of fluxes in
concentrations across the quiet-Sun network
(\citealp{schrijver+etal1996e}, for a global 2D model see
\citealp{2016ApJ...830..160M}). There, the emergence
of ephemeral regions, advection into the supergranular boundaries, merging,
fragmentations, and cancelations maintain a wide distribution of fluxes,
rapidly decreasing in number with increasing flux content. I~called the
process magnetochemistry and Alan Title named the resulting pattern the
magnetic carpet (Figure~\ref{fig:carpet}). \cite{simon+etal2001} followed
up on this with ``cork movies'' in a two-dimensional flow pattern mimicking
supergranulation to show how the ephemeral-region population is crucial
in maintaining the mixed-polarity network.

I should emphasize that the ephemeral-region population has a much weaker
dependence on cycle phase than the active-region population. In fact, it
appears that there is a background population in the least-active areas
of the quiet Sun that is essentially independent of the cycle, as borne
out by the dedicated work of \cite{livingston+etal2007}. They observed
a set of Fraunhofer lines over 33 years, including the Ca\,{\textsc{II}} K
line. When compiling observations of the quietest disk-center measurements
(in a 2' circular patch) over each of the many successive solar rotations,
they found a cycle-independent level: ``Our center disk results show that
both Ca\,{\textsc{II}} K and C\,I 5380 \AA \ intensities are constant, indicating
that the basal quiet atmosphere is unaffected by cycle magnetism within
our observational error.'' Such remarkable dedication to measurements (by
the same people with the same instrumentation) can reveal important clues,
in this case of impact even, e.g., for studies of solar irradiance on terrestrial
climate \citep{2020GeoRL..4790243Y}. Note, by the way, that the use of
the term ``basal'' by \cite{livingston+etal2007} --~a minimum average level
of magnetism of the quiet-Sun network~-- is distinct in its definition
from what I discussed above; see footnote~\ref{foot:basal}.

In addition to an extended temporal base, we can also benefit from an extended
spatial range. As an example in the context of the magnetic carpet, I mention
\citet{thornton+parnell2010}. They identify clumps of photospheric flux
from $10^{16}$ Mx up to $10^{23}$ Mx, from internetwork scales observed
with \textup{Hinode}'s Solar Optical Telescope up to active-region scales
observed by \textup{SDO}'s HMI. They find a smooth number distribution over
18 orders of magnitude in frequency
\citep[see a study by][on the shape of
the distribution]{2023ApJ...943...10S}. Note, however, that there is a
pronounced cycle modulation at the large scales in contrast to what happens
at small scales, as is the case for the emergence frequency of bipolar
regions (Figure~\ref{fig:powerlaws} \emph{left}). Another example of a lesson
from an extended spatial range is discussed in Section~\ref{sec:corona}
(and shown in Figure~\ref{fig:powerlaws} \emph{right}).

\begin{figure*}
  \centerline{\includegraphics[width=0.9\textwidth,trim=0mm 150mm 0mm
     0mm, clip=]{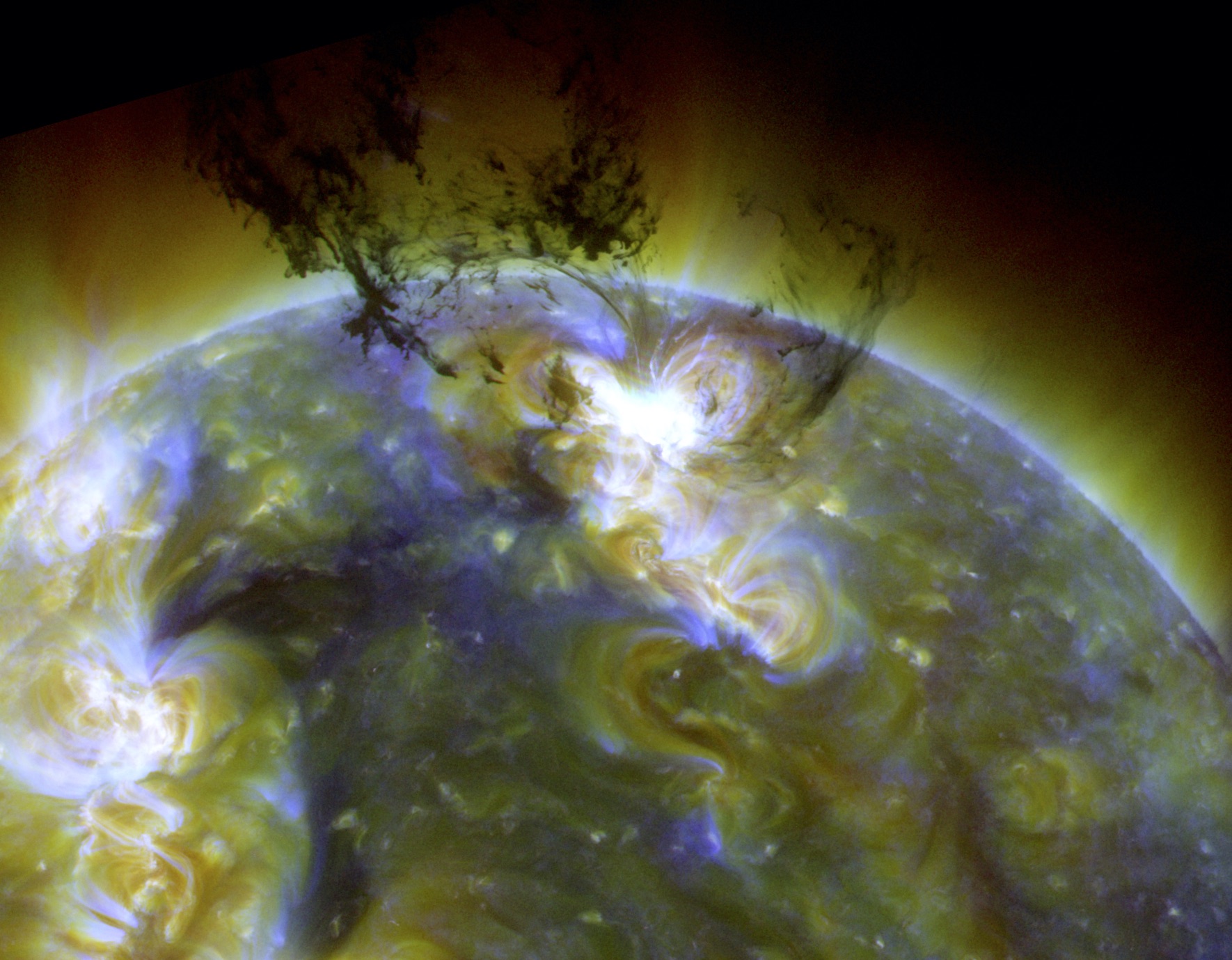}
              }
 \caption{The Sun has provided us with prototypes of many of the phenomena
that we consider responsible for stellar magnetic variability, but Fabio
Reale and colleagues took it one step further when they saw this: SOL2011-06-07T06:16
(here seen in a 171-193-211 \AA \ \textup{SDO}/AIA composite rotated by
$+90^{\circ}$) remains, as far as I know, the most impressive eruption
of relatively cool matter from the Sun observed by \textup{SDO}. The return
of much of that matter, in what struck me as a gigantic bombing raid on
the chromosphere, looked to Fabio's group as a plausible solar counterpart
to what happens in stellar accretion
\citep{2013Sci...341..251R, 2019ASSP...55...43O}.}
\label{fig:accretion}
\end{figure*}
%

\subsection{Stellar Magnetic Activity}
\label{sec5.5}

In this summary, I am skipping multiple topics that I worked on in the
course of these years: cool-star coronal spectroscopy with \textup{EXOSAT},
\textup{EUVE}, \textup{ASCA}, and (preparatory work for) \textup{AXAF}; the deviation
from flux--flux relationships of mid to late M-type dwarf stars; the suppressed
activity of F-type main-sequence stars and the enhanced activity of tidally
interacting binaries given their rotation period; the exploration of stellar
coronal loop geometries and the possibility of nonnegligible optical depth
in the strongest coronal lines; rotation--activity relationships;
{\ldots}  I do not want these stellar topics to make this Sun-centered memoir
unduly long; they could fill a book. In fact, they do. In the early 1990s,
I had been pushing Kees Zwaan hoping to convince him to write a book about
solar and stellar magnetic activity, a field in which he was one of the
pioneers. When he finally agreed, it was on the condition that
I would be his coauthor \citep{schrijver+zwaan99}. This sparked a period
of some three years in which we would both write on our respective sides
of the world, occasionally coming together for a week of intense discussions
in the beautiful, quiet forest-setting of Kees' home in Driebergen in the
center of the Netherlands or, when Kees and his wife Prisca were visiting
Iris and me, in our home in San Jose, CA. Writing a book, I quickly realized,
is hard but rewarding work. I~learned so much from processing and integrating
the literature and from discussing things with Kees.

Leaving so much stellar work out of this memoir goes against my conviction
that we can only learn how the Sun's magnetism has affected, affects, and
will affect Earth by using the population of cool stars: looking at a
mix of young and old stars, stars with different fundamental properties,
with different abundances, and with or without tidally interacting neighbors
is critical to advancing solar physics, as this essentially provides us
with equivalents of both a time machine and an astrophysical laboratory
(cf., Figure~\ref{fig:accretion}). Of course, there are interfaces at meetings
where solar and stellar astrophysicists come together, and a few of us
actually work in these two seemingly distinct areas, but funding agencies
are remarkably efficient at keeping these closely related and mutually
dependent specialties apart. Moreover, much to my (naive?) surprise, these
two communities often see each other as competitors for resources. I~remember
a discussion at an AGU meeting where I suggested that a better integration
of cool-star science with the solar-heliospheric community would benefit
all, only to be rebuffed brusquely by the statement that ``we're not giving
\emph{them} any money.'' It seems the tide is turning, though: with the
discovery of exoplanets and the growth of the field of planetary habitability,
the ``solar--stellar connection'' is evolving into an all-encompassing
stellar--astrospheric--planetary ecosystem science
\citep[e.g.,][]{2023FrASS..1064076G}, i.e., into a field that could be
referred to as comparative heliophysics --~a point to which I return in
Section~\ref{sec:heliophysics}.

\section{Emergence, Dispersal, and Cancelation of Magnetic Field}
\label{sec:dispersal}

\subsection{The Plage State and Flux Dispersal}
\label{sec6.1}

The development of the second-generation construction kit to provide deeper
insight into the connection between local and global flux--flux relationships
through the distribution ${h}_{t}(|\phi |)$ of flux on the solar/stellar
surface required several enabling steps. First among these was an analysis
of how active regions decay into the surrounding network. The prevailing
idea behind this was, and remains, that it is a consequence of the random
walk of flux concentrations caught in the evolving surface convection,
dominated by supergranulation
(initially proposed
by \citealp{leighton64}, although he missed the role of meridional advection on the largest
scales of the process that was not picked up on
until the Ph.D. thesis by \citealp{Mosher1977}, a text that
greatly helped shape my understanding of flux transport). However,
such a decay should lead to a gradual transition from high flux-density
interior regions into low flux-density network, with a profile gradually
spreading out while reducing its amplitude until the active region has
disappeared into the network. That concept clashed with my finding that
active regions have a common mean flux density of some 100 Mx\,cm$^{-2}$
within a fairly well-defined periphery \citep{ref128}. When Alan Title
learned about this so-called plage state via Kees Zwaan, he invited me
for an extended visit to Palo Alto, temporarily taking me away from my
postdoc position in Boulder.

Alan Title, Dick Shine, George Simon, and a team of colleagues had been
analyzing observations of the solar surface taken during a space shuttle
flight
\citep[with the Solar Optical Universal Polarimeter (SOUP)
instrument on \textup{Spacelab\,2} that recorded images from the 30-cm telescope
onto 35~mm film that was later
digitized;][]{1986AdSpR...6h.253T,1988ApJ...327..964S}. They had developed
a flow-mapping algorithm (``local correlation tracking'') to map the
divergence of the flows, and then trace the motions of simulated test particles
in that flow \citep{1988ApJ...327..964S}. They called the resulting image
sequences ``cork movies.'' Alan showed me these and suggested they might
hold a clue to the plage state.

I was given access to the original data and the derived products, stored
on what was then state-of-the-art: so-called optical disks that enabled
quick data access and efficient viewing. I~spent many hours staring at
these movies. I~cannot reconstruct whether it was Alan Title or Kees Zwaan
who made the suggestion, but as I was visiting Palo Alto, I connected with
Sara Martin at BBSO/CalTech to see if they had suitable observations of
active regions. After reviewing many hours of films in the Pasadena archive
--~trying to find image sequences with the least interruptions by weather
and distortion by atmospheric turbulence~-- we did identify the kind of
observations we were looking for, and tracked the movement of flux concentrations
over a five-day period.

The combination of these space- and ground-based data led me to the conclusion
that perhaps a reduction of supergranular scale size (possibly through
a suppression of the largest scales) combined with flux movements primarily
along the boundaries between convective cells could account for the plage
state and for a contrast between effective flux diffusion coefficients
of more than a factor of two between inside and outside active regions.
Some time later, I was privileged as a largely unknown postdoc to sit down
with a very kind Gene Parker over a lunch during a science meeting in Freiburg,
Germany, to discuss possible reasons for this, but even using back-of-the-envelope
(well, actually front-of-a-paper-napkin) sketches, we came no closer to
a solution
\citep[not surprisingly: we still
do not fully understand supergranulation, see, e.g.,][]{2018LRSP...15....6R}.

Later work with then-Ph.D. student Mandy Hagenaar on \textup{SoHO}/MDI
magnetogram sequences suggested that a flux dispersal coefficient
$D$ useful for the modeling that I was thinking of would depend on the
flux $|\Phi |$ of magnetic concentrations. In quiet, mixed-polarity regions,
flux concentrations would typically be small, being fragmented in the flows
and weakened by cancelation against opposite polarity, and these small
flux concentrations would be quite mobile, corresponding to a relatively
high diffusion coefficient. In the predominantly unipolar active-region
plage, frequent mergers and the absence of cancelation collisions results
in larger concentrations with lower mobility and thus reduced effective
diffusion. That nonlinearity in the dispersal of flux through a flux-dependent
diffusion coefficient proved an important ingredient in the surface flux-dispersal
modeling that formed the foundation of the second-generation construction
kit. Note that \cite{worden+harvey2001} independently confirmed the need
to lower the flux dispersal within magnetic plages based on full-sphere
flux transport modeling (implemented in different ways in the LMSAL assimilation
model\footnote{\url{https://www.lmsal.com/forecast/}.\label{fn:vsl}}
and in the ADAPT model described by \citealp{2015SoPh..290.1105H}).

Surface flux dispersal was developed by Ken
\cite{1972SoPh...26..283S}, Neil Sheeley, Yi-Ming Wang, and colleagues
\citep[see][]{2005LRSP....2....5S}. Until 2000, this modeling was essentially
in the hands of Wang and Sheeley, who, through a series of illustrative
and insightful papers, showed the fundamental applicability of the concept
that the Sun's magnetic field, once at the surface, behaved essentially
as a freely advecting signed scalar subject to diffusive dispersal while
embedded in the large-scale differential rotation and meridional advection.
However, I wanted to develop a model that could be used as a numerical
laboratory, in which each of the components could be varied in order to
explore conditions on other real or hypothetical stars as well as for the
Sun in time \citep{schrijver2000}. Others have done so too
\citep[most using a
continuum description for the field, in contrast to the gridless particle-tracking
approximation that I used, or the grid-warping concept independently developed in
parallel to my work by][]{worden+harvey2001}, demonstrating the value of
different approaches and assumptions
\cite[e.g.,][]{mackay+etal2002,2014ApJ...780....5U,2016AnA...586A..73M,2023AnA...672A.138N,2023arXiv230301209Y},
and with different modes of data assimilation
\citep[see a comparative
assessment by][]{2023ApJ...946..105B}. The implication behind that statement
deserves an explicit formulation: funding agencies often consider only
a single model needed, but I have repeatedly seen the value of having
several independently developed models in homing in on a deepened understanding
of a process; other examples can be found in Section~\ref{sec:appl} on
coronal heating and Section~\ref{sec:nlfff} on coronal field modeling.

In the first application of my flux-dispersal model, I demonstrated that
the time-dependent distribution function of flux densities across the solar surface,
${h}_{t}(|\phi |)$, arises as a natural consequence of the flux emergence
patterns (including the butterfly diagram, nesting properties, ephemeral
regions and active regions, and all the other ingredients based on solar
observations) through the solar cycle. I~also showed that by only changing
the total number of emerging bipolar regions, local and disk-integrated
flux--flux relationships naturally transformed into each other (through
Equation~\ref{eq:fftrans}), even when going well beyond the flux-emergence
rates exhibited in the course of the modern-day solar cycle, thus showing
that a truly Sun-like star with a range of different levels of activity
would be consistent with many stellar observations. There is a problem
of nonuniqueness here, of course; the above does not mean that the dynamo
in young, active stars does not behave differently from that of the Sun,
but it implies it need not do so.

\begin{figure*}
  \centerline{\includegraphics[width=0.9\textwidth,trim=0mm 94.5mm 00mm 89.6mm,clip=]{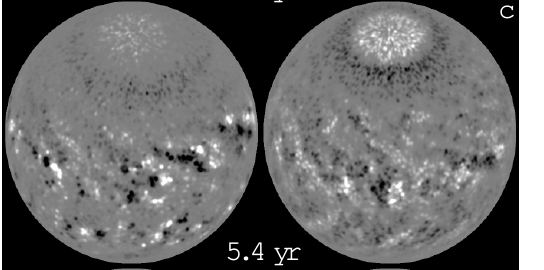}
              }
\caption{Simulated magnetograms for the Sun (\emph{left}) and a Sun-like
star (or a young Sun) with a strongly enhanced rate of flux emergence (\emph{right}),
visualized from a perspective 30$^{\circ}$ above the stellar equator about
midway through a magnetic cycle. The observed decrease of flux dispersal
with increasing size of the concentrations (see Section~\protect\ref{sec:appl})
suggests that strong polar caps, if not polar starspots, may form on more
active stars through increased clustering of flux that is swept poleward
in the meridional advection, even if the geometry of the butterfly diagram
does not change. (From \citealp{schryver+title2001}.)}
\label{fig:PolarSpots}
\end{figure*}
%

\subsection{Select Applications of the Second-Generation Construction Kit}
\label{sec:appl}

A large increase in the rate of flux emergence to mimic the most active
stars, led to an unanticipated outcome through the functional dependence
of $D(|\Phi |)$: in the most active simulated stars, the mean field strength
of the polar caps could match or exceed that found in the penumbrae of
sunspots, thus suggesting one possible formation mechanism of polar starspots
\citep[][see
Figure~\ref{fig:PolarSpots}]{schryver+title2001} that are inferred to exist
on such stars through (Zeeman) Doppler imaging. Of course, there
are alternative formation mechanisms. In rapidly rotating stars, flux emergence
is likely affected by Coriolis forces that would act to deflect rising
flux to higher latitudes; the effects of that for flux distribution on
the stellar surface and the potential formation of polar spots have been
studied by, e.g., \cite{holzwarth+etal2007} and
\cite{2018A&A...620A.177I}.

Another surprise resulted when Marc DeRosa, Alan Title, and I drove the
model with a sunspot record that mimicked that observed for the Sun since
the Maunder Minimum. We found that big differences between successive cycles
could cause the simulated polar caps to occasionally not reverse polarity,
inconsistent with historical records \citep{schrijver+etal2002a}. We
proposed a flux-decay process that was later given a physical foundation
by \cite{baumann+etal2006}. \cite{wang+etal2002c} suggested alternatively
that modulation of the meridional advection could (also) be involved in
avoiding such nonreversals of the polar-field polarity, while
\cite{cameron_etal2010} looked into the effects of changing active-region
tilt angles. How much any of these processes are involved in the solar
dynamo remains a field of study
\citep[e.g.,][]{2014SSRv..186..491J,2023arXiv230301209Y}.

\begin{figure*}
 \centerline{\includegraphics[width=0.5\textwidth,trim=58mm 58mm 0mm
    0mm,clip=]{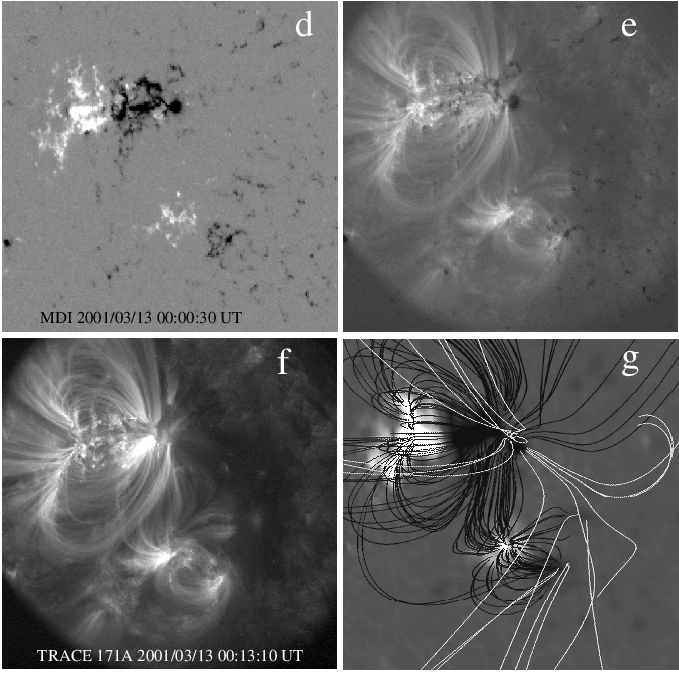}
                     \includegraphics[width=0.5\textwidth,trim=58mm
                     0mm 0mm 58mm, clip=]{AR010313a.eps}
              }
\caption{The heliospheric field has its roots not only in large-scale coronal
holes over quiet Sun, but also in active regions, often involving sunspots,
as in this example. \emph{Left}: a blend of \textup{TRACE} 171 \AA \ and a
\textup{SoHO}/MDI magnetogram shows both coronal loops and sunspots. \emph{Right}:
A global PFSS model based on a low-resolution magnetic map (shown in the
background) shows connections of closed (black) field lines and open (white)
connections to the heliosphere. (Panels from a figure by
\citealp{schrijver+derosa2002b}.)}
\label{fig:sunhelio}
\end{figure*}
After these and other experiments, Marc DeRosa and I transformed the code
so that it could assimilate magnetograms (first those of \textup{SoHO}/MDI,
later \textup{SDO}/HMI) and coupled it to a potential-field source-surface
(PFSS) model that Marc coded up (and that has found frequent use after
Marc made it part of the SolarSoft library). From it we learned
\citep{schrijver+derosa2002b}, for example, how open field originates in
both quiet Sun and active regions, even reaching upward from sunspots (Figure~\ref{fig:sunhelio}).
The flux-assimilation-and-transport model (see footnote~\ref{fn:vsl}),
along with its PFSS superstructure, continues to run at LMSAL, at the time
of this writing extending over 28 years, having provided guidance and field
models to many community studies of solar and heliospheric activity.

With the PFSS package came the possibility of exploring coronal heating,
not for individual loops --~which are hard to disentangle from others and
the background diffuse emission in the transparent corona and hard to trace
completely given their multithermal nature~-- but instead for ensembles,
i.e., for entire active regions or, as we attempted, for the global corona.
At my initiative, Markus Aschwanden developed a set of analytical approximations
to quasihydrostatic loop atmospheres with different heating scale heights
and expansions with height \citep{aschwanden+schrijver2002} that enabled
us to quickly populate multitudes of coronal ``loops'' that were traces
of the field from the PFSS model. A~student, Annie Sandman, then bravely
took on the challenge of populating a volume with such loop atmospheres
for subsequent 2D rendering. When JPL's Paulett Liewer learned of this,
she asked for examples rendered from different pairs of viewing angles
as they were preparing for the launch of the \textup{STEREO} spacecraft in
October of 2006. When we put these images on our (then brand new) stereoscopic
displays, one of the things we learned was that the human brain is best
at interpreting these transparent volumes as 3D structures when the separation
in viewing angles is of the order of 5\,--\,10 degrees; that is not surprising,
as that is what we use in daily life, too. However, it makes one think
of a mission that would provide such image pairs to interpret coronal structures
and the initiation of CMEs, which would be of great help since we cannot
at present reliably model the Sun's nonpotential fields; more on that in
Section~\ref{sec:flares}.

With this visualization code in hand, we experimented with different parameterizations
of the dependence of coronal heating on the base field strength and loop
length \citep{schrijver+etal2004}. Simulated images for the EUV and soft
X-ray corona could approximate observations for a heating function that
pointed us toward a mechanism dominated by nanoflare reconnection induced
by footpoint braiding, at least for loops with lengths up to some 50{,}000~km.
For longer loops (typically the cooler ones), however, more heating was
needed than provided by that best-fit scaling of the coronal heating, so
it remained an ambiguous outcome and a limited success.

Later, \cite{warren+winebarger2007} performed a similar visual rendering,
but focused on a single active region. They based their modeling on hydrodynamic
simulations of loops in a PFSS model rather than approximated quasistatic
ones, finding ``that it is possible to reproduce the total observed soft
X-ray emission in all of the [\textup{Yohkoh}] SXT filters with a dynamical
heating model, indicating that nanoflare heating is consistent with the
observational properties of the high-temperature solar corona.'' They,
too, noted problems in describing the cooler EUV loops.

\cite{lundquist+etal2008} chose yet another approach: for a set of 10 active
regions, they computed a nonpotential ``quasi force-free'' field model
and populated loops with quasistatic atmospheres. They, too, homed in on
braiding-driven DC heating, but concluded that although ``this parameterization
best matches the observations, it does not match well enough to make a
definitive statement as to the nature of coronal heating. Instead, we conclude
that (1) the technique requires better magnetic field measurement and extrapolation
techniques than currently available, and (2) forward-modeling methods that
incorporate properties of transiently heated loops are necessary to make
a more conclusive statement about coronal heating mechanisms.'' I could
not agree more with that assessment, even now, 15 years later. Moreover,
the views that a single heating mechanism dominates throughout the entire
corona and that different energy-conversion mechanisms operate independently
of each other are increasingly questioned
\citep[see for example
the reviews by][]{2006SoPh..234...41K,2015RSPTA.37340269D}.

I will mention one more application of the surface flux-dispersal code,
with some introduction. With the increasing number of transiting exoplanets
came the need to also understand the contributions to the transit light
curves from structures on the surfaces of their parent stars, particularly
when transit spectroscopy is used to derive information about the exoplanetary
atmosphere
\citep[{e.g.},][]{2018ApJ...853..122R,2019AJ....157...96R}. Conversely,
the transiting exoplanet can tell us about the structures on stellar
surfaces down to resolutions that are far below what can be achieved by
rotational-modulation or Doppler-imaging studies
\citep[e.g.,][]{2011AnA...529A..36S}. To explore that for myself, I used
the construction kit \citep{2020ApJ...890..121S} to simulate the quiet
regions, active plage, and starspots on stars with flux emergence rates
up to 30 times the solar-maximum value, and used the results of the MURam
code for facular contrast versus limb distance at three wavelengths published
by \cite{2017A&A...605A..45N}. The conclusions exemplify the value of the
solar paradigm to astrophysics just as much as the value of stellar studies
to solar physics: ``This (1) indicates that the solar paradigm is consistent
with transit observations for stars throughout the activity range explored,
provided that infrequent large active regions with fluxes up to
$\sim 3 \times 10^{23}$ Mx are included in the emergence spectrum, (2)
quantitatively confirms that for such a model, faculae brighten relatively
inactive stars while starspots dim more-active stars, and suggests (3)
that large starspots inferred from transits of active stars are consistent
with clusters of more compact spots seen in the model runs, (4) that wavelength-dependent
transit-depth effects caused by stellar magnetic activity for the range
of activity and the planetary diameter studied here can introduce apparent
changes in the inferred exoplanetary radii across wavelengths from a few
hundred to a few thousand kilometers, increasing with activity, and (5)
that activity-modulated distortions of broadband stellar radiance across
the visible to near-IR spectrum can reach several percent.'' I note that
the apparent need to include some very large active regions in the most
active stars may be an alternative formulation to the need for enhanced
active-region nesting to explain the observed rotational modulation of
stars \citep{2023AnA...672A.138N}.

This last project led to an accepted proposal to ISSI, together with Louise
Harra, Kevin Heng, Moira Jardine, Adam Kowalski and Ben Rackham, to bring
together the solar, cool-star, and exoplanetary communities in a meeting
in Bern to learn how to disentangle the various contributions to transit
spectra to derive information on stellar magnetism and on exoplanetary
atmospheres. Unfortunately, we had to cancel that meeting because of both
COVID-19 and personal health reasons. It would have been a truly fun project!
In this context, let me point out the excellent book by Jeff
\cite{2019LNP...955.....L} the title of which states its focus: ``Host
stars and their effects on exoplanet atmospheres.''

\section{Coronal Structure and Dynamics}
\label{sec:flares}

\subsection{Learning from the Corona}
\label{sec:corona}

By the late 1960s, rocket-borne X-ray telescopes had already revealed that
the solar corona is structured in loop-like shapes determined by the Sun's
magnetic field. A~decade later, \cite{ref271} developed their famous quasistatic
loop model linking loop length, density, temperature, and heating rate
through scaling laws. When I started working on the interpretation of stellar
XUV spectra, it was in the context of this ``RTV'' loop concept. In the
1980s, many authors assumed that coronal loops were long-lived structures
and consequently felt the need to prove that loops would be thermally stable
when subjected to variable heating. I~even started writing a review of
such studies that I thought might be included in my Ph.D. thesis. I~never
completed that chapter, because many of the studies appeared to
rely on rather artificial assumptions and approximations. It was not until
the late 1990s, however, that the \textup{TRACE} mission with its arcsecond
resolution (slightly better even than that of \textup{SDO}'s AIA) and 30\,--\,60-s
image cadence demonstrated that coronal loops were not long lived at all.

\begin{figure*}
   \centerline{\includegraphics[width=0.96\textwidth,,trim=0mm
       80mm 0mm 80mm,clip=]{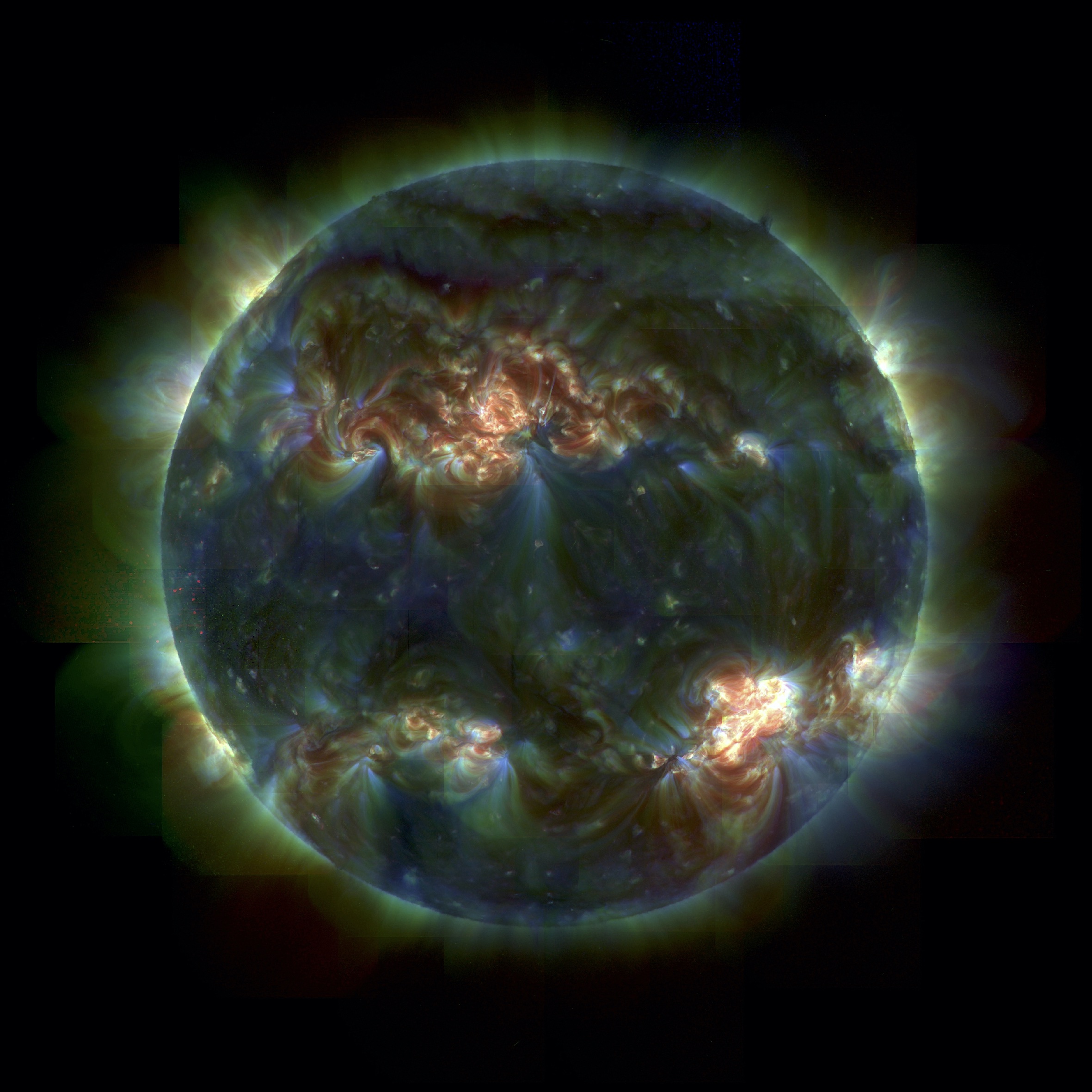}
              }
\caption{A three-wavelength full-disk mosaic of the EUV corona observed
by \textup{TRACE} on 1999/08/02. These images (put together with software
developed by Ed DeLuca at CfA, Sam Freeland at LMSAL, and others) took
well over a full \textup{TRACE} orbit of 96 min, during which Earth eclipses
and high-latitude radiation zones had to be avoided. \textup{SDO}/AIA made
a gigantic leap forward by providing eight-wavelength full-Sun images on
a 12-s cadence with very few interruptions throughout its now 13 years
of service.}
\label{fig:TRACEmosaic}
\end{figure*}
\textup{Yohkoh}'s X-ray telescope and \textup{SoHO}'s Extreme-ultraviolet Imaging
Telescope (EIT) had been observing the Sun's corona for some seven and
three years, respectively, by the time \textup{TRACE} was launched. Their
combined full-Sun, multithermal coverage provided many new insights into
the large-scale physics of the corona
\citep[{e.g.},][]{1998Ap&SS.261...81P}. When \textup{TRACE} came along
in 1998 (Figure~\ref{fig:TRACEmosaic}), it opened a new window on the smaller
length and time scales. Its initial discoveries were formulated in a paper
that appeared only a year after its science observing began
\citep{schryver+etal99a}.

The first year of \textup{TRACE} was a very busy time for the science team
spread out over LMSAL, CfA, and Montana State University (MSU). The planners
rotated on a weekly schedule, flying out to Goddard Space Flight Center
(GSFC) in Greenbelt, MD, once every four to six weeks, to spend much of
a week holed up in the basement of the flight operations building, one
floor below the much more spacious \textup{SoHO} counterpart. We coordinated
our planning with \textup{SoHO}'s instruments, thus increasing the coverage
in spectral, temporal, and spatial dimensions (although, as I mentioned
above, \textup{SoHO}'s instruments were often off looking at different solar
targets forcing us to choose who to coordinate with). The mornings were
taken up by formulating an observing plan, selecting a target (although
after the first months of exploratory observing we generally stuck with
a given target for multiple days), and communicating with the \textup{SoHO}
team and members of the scientific community with specific observing requests,
often in coordination with ground-based observatories around the globe.

Afternoons and evenings were dedicated to reviewing the observations of
the preceding day(s) (using a movie-generation toolset developed by Dick
Shine) and distilling out of these what we could learn about the solar
corona. Among other things, we noted the shifting around of sites of strong
coronal heating, frequent coronal rain in apparently quiescent settings,
pulsations traveling into loops upward from the solar surface, linkages
between distant regions with near-simultaneous emission features at times
of flares and eruptions, and the intricate dynamics of the transition region
seen in \textup{TRACE} imagery as ``moss.'' Many of these early discoveries
spurred series of subsequent studies. For example, the transverse oscillations
that excited Dick Shine, Charles Kankelborg, and me as we first saw them
upon reviewing a preceding day's \textup{TRACE} sequence, led to hundreds
of studies on transverse coronal-loop oscillations: their initial study
reported by \cite{aschwanden+etal99} has more than 750 citations in ADS.

For twelve years, I collected images and movies by \textup{TRACE} from our
daily inspections by the planners and sometimes from the community to put
online as ``picture of the day'' (POD). After a while, they were published
weekly, a practice kept up throughout \textup{TRACE}'s 12-year operational
period. Alan Title (the PI for most of the mission's duration) and I (science
lead and then PI for the last four years) shared the same philosophy: there
is so much to discover in what has been publicly funded, that we wanted
to present events of interest to the worldwide community, hoping they would
pick up on their analysis and interpretation. In many cases, it worked;
although we never did a quantitative assessment, we have the impression
that many of our PODs were the seeds for refereed publications.

This brings me to a note about the ``open data policy'' that is now mandatory
within NASA's Science Mission Directorate. Although Earth science data
at NASA were openly accessible since 1994, there was no such policy in
the other science divisions. Alan Title pushed hard for this in the Sun-Earth
Connections Division on the \textup{TRACE} mission. Let me quote him directly
on that (from an AIP Oral History interview): ``When we flew \textup{TRACE},
I decided that all the data that came out of \textup{TRACE} would be available
to everybody at the same time as we got it. There was a lot of internal
discussion and a lot more discussion with our co-investigators. All kinds
of concerns, for example, what would happen if after you've worked on a
project eight or nine years and because the observations are public a person
who was not on the team makes an interesting discovery and publishes? Why
can't we have, as had been done in the past, the right to review any paper
that comes out from the data? Or, why can't we have this data stay in a
locked box in our building for a year before anybody else gets to look
at it or forever before anybody gets it? [\ldots] there's a moral issue
too. The government spends 100 million dollars or more of the nation's
science budget on your instrument, and it costs another large amount of
money to launch the mission. It is my opinion that you really have an obligation
to society that this knowledge is used as broadly and as well as possible.
The other thing is, that no matter how clever you are, there's somebody
else who will have new insights. [\ldots] The \textup{TRACE} policy worked
really well. We didn't have any of the problems that were so worrisome
initially. It got NASA to agree that this was the way to go. [\ldots]
The open data policy was the hardest battle that we fought.''

When preparing for \textup{SDO}, we settled on a different model to bring
solar events to the community's attention. Phil Scherrer, the \textup{SDO}/HMI
PI, had formulated the idea that with a full-Sun observatory with continuous
observing, the community would be operating in a mode in which they would,
in effect, be observing the archive. In parallel, Neal Hurlburt, our data
system lead, came up with the idea of an event database: instead of having
to download and search through terabytes of HMI and AIA data to find events,
Neal's programmers developed an archive of events to be populated by team
members who would identify events in observations on a daily basis, complemented
by software that would run on summary data in house or elsewhere in the
world. That led to the heliophysics events knowledgebase
\citep[HEK,][]{aiadata}\footnote{\url{https://www.lmsal.com/hek/}.} and a search
interface for users (iSolSearch)\footnote{\url{https://www.lmsal.com/isolsearch}.}
which now has a web browser interface and also IDL- and Sunpy-based AIPs.
This, too, has guided the solar-physics and heliospheric communities to
events of interest that led to research publications, although --~as I
noted above~-- it still too infrequently leads to population studies.

The \textup{TRACE} team executed its own research, too, of course. I~focused
preferentially on things where being right on top of the data archive of
\textup{TRACE} and near to that of \textup{SoHO}'s MDI was beneficial to reviewing
large volumes of data. These included the study of transverse loop oscillations
compiled from years of \textup{TRACE} observations, an assessment of the conditions
under which the corona over active regions showed clear signs of nonpotentiality
(using the flux assimilation and PFSS modeling described above), and --~in
2010, the final year of \textup{TRACE} operations~-- a study that showed that
the distribution of the number versus the size of eruptions from ephemeral
regions smoothly extended such a distribution for coronal mass ejections
(see Figure~\ref{fig:powerlaws}).

\subsection{Nonpotential Fields and Its Instabilities}
\label{sec:nlfff}

For the first two decades of my scientific career, I steered clear of solar
and stellar flares and coronal mass ejections, mostly because I saw them
as infrequent interruptions of the (relatively) quiescent background state,
and also because it seemed that observational data rarely sufficed for
a comprehensive picture. I~avoided flare studies despite working at the
space research laboratory that built the Hard X-ray Imaging Spectrometer
(HXIS) for the \textup{Solar Maximum Mission} that was launched in 1980, just
prior to my Ph.D. phase. However, the frequently beautiful \textup{TRACE}
observations of eruptive and explosive processes made a huge difference
in my view of flares. Then, NASA's Sun-Earth Connections (SEC) Division
launched its ``Living With a Star'' (LWS) program (a term that, I think,
originated from Bill Wagner at NASA/HQ) and I found myself invited to its
founding LWS Science Architecture Team; more on that in Section~\ref{sec:heliophysics}.
When we were awarded the \textup{SDO}/AIA contract in 2003 as part of the
first LWS mission, I could no longer walk around solar energetic events
or its causes at and below the solar surface. Also, when the SEC Division
became the Heliophysics Division in 2007 (a name adopted by Division director
Dick Fisher when the NASA administrator challenged him more or less on
the spot to come up with a one-word name for the division ``that had better
end in -physics''), it was clear that flares and eruptions needed to be
an integral part of the \textup{SDO}/AIA science program, which I helped shape
as the team's science lead. What I mostly focused on, though, was not so
much the flare or eruption itself as the conditions under which nonpotential
magnetic energy came to exist in the corona and the circumstances that
might trigger the conversion of that free energy into other forms of energy.

\begin{figure*}
  \centerline{\includegraphics[width=0.75\textwidth,clip=]{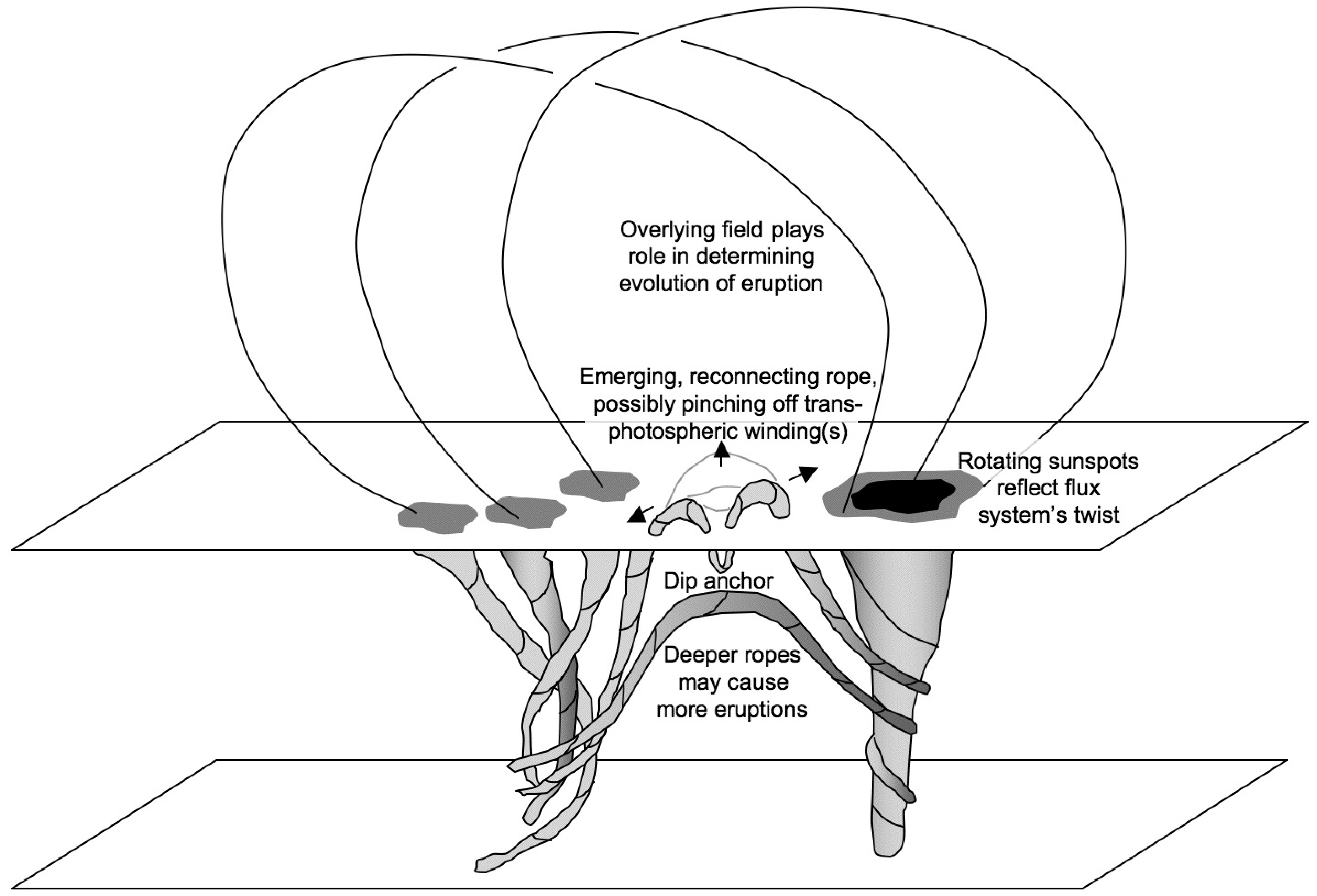}
              }
\caption{Schematic representation for an emerging field configuration involved
in flaring or flux-rope eruption. (From \citealp{schrijver2009b}.)}
\label{fig:flarefluxmodel}
\end{figure*}

Marc DeRosa, Alan Title, and Tom Metcalf (who died far too young in a skiing
accident in the Rocky Mountains) joined me on the path to magnetic instabilities
by reviewing the configurations and dynamics of the photospheric field
that were associated with substantial nonpotentiality of the coronal field
of active regions as evidenced by mismatches between observed loop patterns
and modeled potential fields. Then, there was an opportunity that came
out of a workshop to analyze observations of a large X17 flare observed
from gamma rays to optical with US and Russian spacecraft. Another step
to studying eruptions came when \textup{TRACE} observed the accelerating rise
of a limb filament (analyzed by Chris Elmore, a summer student, and interpreted
with modeling by Bernhard Kliem and Tibor T\"or\"ok). Also, my initial
learning curve culminated in a review presentation at the COSPAR general
assembly in Montr\'eal, Canada, in 2008 (cf., Figure~\ref{fig:flarefluxmodel}). I~learn far better by doing than by reading, so getting hands-on experience
with such studies and writing a review proved most helpful.

In preparing for the flight phase of \textup{SDO} it seemed a good idea to
assemble an international team of experts to learn to use the future \textup{SDO}/HMI
vector magnetograms to create nonpotential models of active-region coronae.
So, in 2004, Marc DeRosa, Tom Metcalf, and I invited experts in nonpotential
force-free field (NLFFF) modeling to come to LMSAL in Palo Alto to prepare
for \textup{SDO}: we were joined by Yang Liu, Jim McTiernan, St\'ephane
R\'egnier, Gherardo Valori, Mike Wheatland, and Thomas Wiegelmann. In
my unfamiliarity with the problem, I had imagined that the key problems
would be validation of the vector field map (with its intrinsic ambiguity
in the field direction transverse to the line of sight) and acquiring adequate
resources to efficiently compute the field. How wrong I was! That initial
meeting caused the team to walk back from that ambitious modeling goal
and to start with what seemed the simplest first step: reproduce an analytically
known NLFFF model based on the lower-boundary vector field. This led to
a series of meetings that we held both in the USA and in Europe, resulting
in five indepth team studies and several papers by various subgroupings
of the evolving team. The result: it may be possible to successfully
model the nonpotential field of solar active regions based on photospheric
vector field measurements, but one should never trust the result of any
of the methods unless there is a validation of that model by a match to
the observed solar corona \citep{derosa+etal2008}. How to quantify the
quality of such a model still remains to be seen. I~continue to be amazed
that NLFFF models are produced (and let pass by referees) without at least
a visual comparison with coronal images!

In parallel to the NLFFF project, I started looking at the evolving configurations
of active-region magnetic fields involved in major flaring using the multitude
of \textup{SoHO}/MDI magnetograms. After reviewing 2500 magnetograms and almost
300 M and X-class flares, there was no pattern that I could recognize that
was uniquely associated with such large flares. That was hardly surprising
because many colleagues had spent decades looking for such a signature.
However, I did notice ``(1) that large flares, without exception, are associated
with pronounced high-gradient polarity-separation lines" \citep[certainly not a
new finding, see the review by][]{2019LRSP...16....3T}, "while (2) the
free energy that emerges with these fibrils is converted into flare energy
in a broad spectrum of flare magnitudes" \citep[also noted before, as
well as after,
e.g.,][]{2001SoPh..203...87W,2021ApJ...915...38C} "that may well be selected
at random from a power-law distribution up to a maximum value,'' and noted
that such features tend to form in association with flux emergence
\citep[confirmed by][]{welsch+li2008}. Based upon this, I proposed a metric,
which I named ``$R$'', as a measure for the flux encompassed in the area
near high-gradient polarity-separation lines that sets an upper limit to
the flare intensity that could occur \citep{schrijver2007}. That metric
has been used in many subsequent studies as a forecasting metric for flares
or the absence thereof, with \cite{2016ApJ...829...89B} concluding that
in a comparison of metrics ``no participating method [proved] substantially
better than climatological forecasts.'' However, it seems to me that the
final part of the second conclusion above is important because it contains
the possibility that deterministic forecasting of flare magnitudes is,
as a matter of principle, unachievable. Then again, expecting a deterministic
forecast based on a single parameter for a complex, evolving magnetic field
would be naive.

\begin{figure*}
  \centerline{\includegraphics[width=0.9\textwidth,trim=40mm 120mm 60mm
     115mm,clip=]{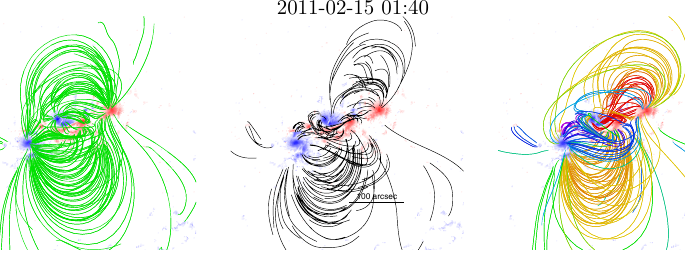}
              }
\caption{Coronal loops can in principle be used to guide nonlinear force-free
field models over active regions. This image composite shows coronal loops
seen over AR 11158, traced from coronal images (\emph{center}), best-fitting
field lines of a potential field (\emph{left}), and a Grad--Rubin solution
(\emph{right}) over the magnetograms (the background red/blue images). The
colors of the field lines in the nonlinear force-free field model in the
\emph{right} panel encode the value of $\alpha $ (defined by
$\nabla \times {\mathbf{B}} \equiv \alpha {\mathbf{B}}$) ranging from strongly positive
(red) to strongly negative (blue) through $\alpha \approx 0$ (green).
(From \citealp{2014ApJ...783..102M}.)}
\label{fig:QGR2014}
\end{figure*}
When Anny Malanushenko joined our group after completing her Ph.D. with Dana Longcope at MSU, I asked her to develop a method of NLFFF modeling
that could be guided by observed loops as a form of scaffolding to which
an algorithm would nudge a model field. She courageously accepted that
challenge and succeeded
\citep[Figure~\ref{fig:QGR2014};][]{2012ApJ...756..153M,2014ApJ...783..102M}!
I hope that her algorithm and others like it help us move forward with
NLFFF modeling.

\subsection{The Long Reach of Magnetism}
\label{sec7.3}

The high-cadence full-Sun view of \textup{SDO}'s AIA revealed connections
between activity in distant regions: filament eruptions or flux emergence
in one location can deform the magnetic configuration sometimes more than
a solar radius away. Combined with the full-Sun field models and \textup{STEREO}
EUVI observations of parts of the Sun not visible from Earth, we could,
in one well-studied case, connect several flares and filament eruptions
to the locations where the field topology was being modified by the emergence
of an active region well behind the solar limb \citep [note that the region's emergence was observed with EUVI, i.e., in coronal emission only, because the \textup{STEREO} spacecraft carried no magnetographs;][]{schrijver+title2010}. A~few years later, we looked into another set of ``sympathetic events''
\citep{2016ApJ...820...16J} using an MHD model to explore mechanisms of
couplings; this became possible after Meng Jin joined us at LMSAL (first
as a Jack Eddy Fellow and eventually taking on the role of PI of \textup{SDO}/AIA
that I had handed over to Mark Cheung when I retired), bringing with him
his expertise with the AWSoM MHD model developed at the University of Michigan.

Paul Higgins and I performed a statistical study of the timings of events
observed by AIA to find that large flares in one region not infrequently
were at least temporally associated with activity in other, distant regions
within a matter of hours, possibly enabling the release of nonpotential
energy there through gradual field deformations or wave-like perturbations,
so that after that release activity in distant regions is temporarily
somewhat less likely. Those effects are relatively weak, not quite
at the 2$\sigma $ level even for the large sample of events that we studied.
\cite{2020SoPh..295...70H} notes, in his Section~5, that hints of such
couplings were already seen in H$\alpha $ flare patrol movies of 1961 that
eventually revealed Moreton--Ramsey waves.

The sensitivity of AIA and its ability to trace the evolution of the corona
through a range of temperatures \citep[e.g.,][]{2015ApJ...807..143C} revealed
that whereas such direct causal links may be relatively infrequent, the
assumption that the coronal field surrounding destabilizing regions is
relatively static is wrong: ``the post-eruption reconfiguration timescale
of the large-scale corona, estimated from the extreme-ultraviolet afterglow,
is on average longer than the mean time between coronal mass ejections
(CMEs), so that many CMEs originate from a corona that is still adjusting
from a previous event''
\citep[][and references
therein]{2013ApJ...773...93S}.

\subsection{With the Help of Solar Meteors}
\label{sec7.4}

Sometimes, opportunities to test our ability to model the Sun's magnetic
field arise from entirely unexpected directions. In the days leading up
to 2011/07/06, \textup{SoHO}'s LASCO coronagraphs observed one of the thousands
of Sun-grazing comets that now populate its archive.\footnote{\url{https://sungrazer.nrl.navy.mil}.}
This one, C/2011 N3 (\textup{SoHO}), was unusually bright in LASCO and could
be traced in the images to the inner edge of the instrument's field of
view. As I was inspecting the AIA data for 2011/07/06, I thought it would
be fun to see if there was any signature of that comet, although I did
not expect anything of that exercise. Much to my surprise, it
did show up: for some 20 min, the comet was observed in up to five of AIA's
EUV channels, first off-limb and later in the corona seen against the disk.
\textup{STEREO-B}'s EUVI also recorded the comet, seeing it off the limb on
closest approach from its near-quadrature position at the time. Orbital
data showed that the comet (likely disintegrating as it sublimated in the
sunlight) penetrated to about 100{,}000~km from the surface before its demise.
AIA data clearly showed the deceleration of the ionized material in the
comet's tail, enabling us to use the anticipated Lorentz force to estimate
the comet's mass-loss rate. A~team of coauthors rapidly assembled around
this first-ever detection of a coronal meteor, which brought me in touch
with a very enthusiastic cometary community that I never anticipated crossing
paths with.

\begin{figure*}
   \centerline{\includegraphics[width=0.75\textwidth,clip=]{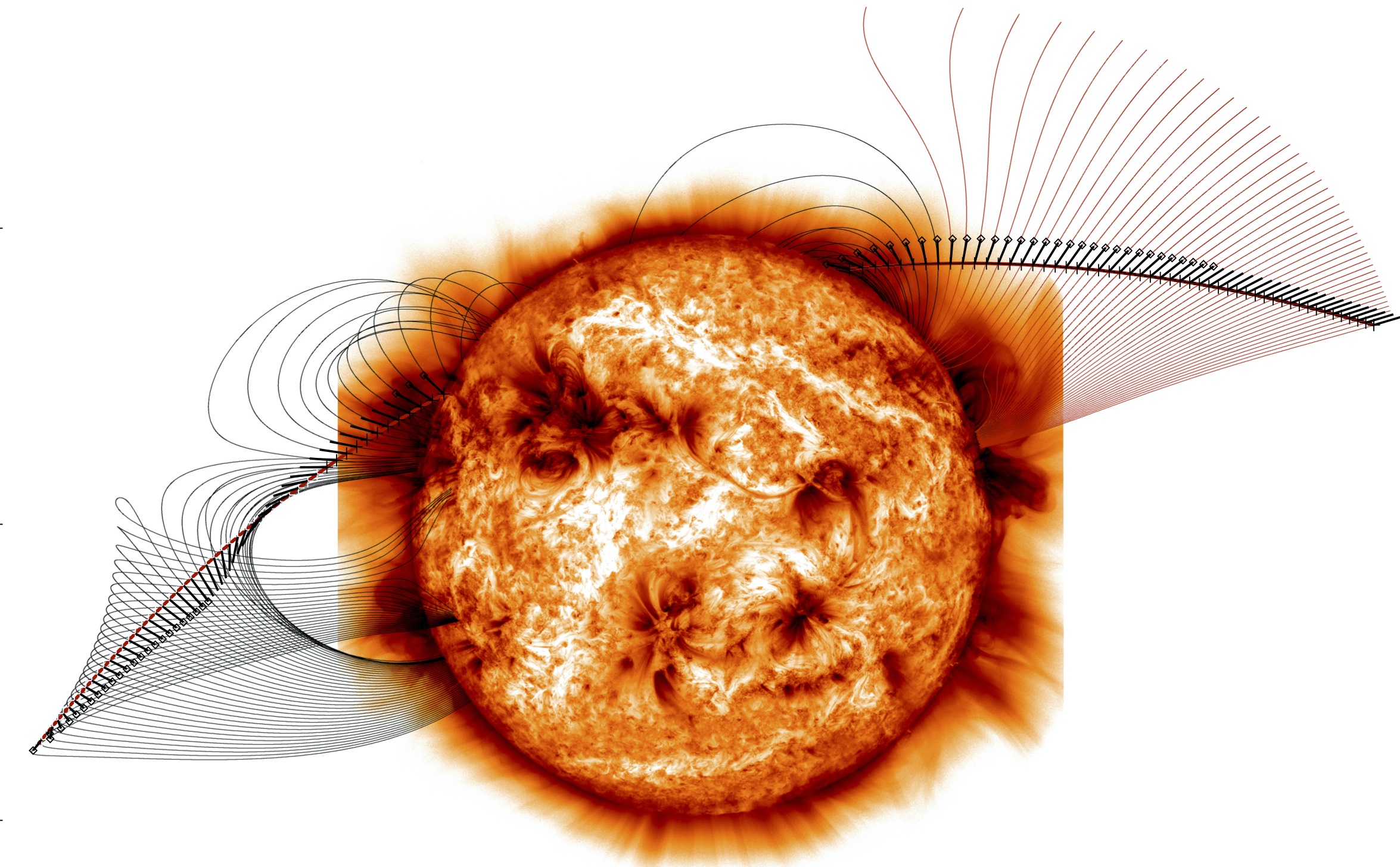}
              }
\caption{A Sun-grazing comet provided a (thus far) unique means to test
MHD models of the connection between the high corona and the heliosphere.
This image shows the (red) orbit of (Kreuz-group) comet C/2011 W3 (Lovejoy)
superposed on a negative of an EUV image of the solar corona. The tangents
(arrows) to the coronal magnetic field projected against the sky were compared
to the trajectories of ionized material escaping from the comet and locking
onto the magnetic field as seen in \textup{SDO}/AIA EUV off-point images (not
shown) as explored by \cite{lovejoymhd2013}. (Courtesy Marc DeRosa.)}
\label{fig:Lovejoy}
\end{figure*}

It made all the teams involved more aware of the potential of detecting
more such Sun-grazing comets, and for some time busy email traffic ensued
each time LASCO detected a promising candidate. Amazingly, we had to wait
for only a few months: on 2011/12/15--16 comet C/2011 W3 (Lovejoy) obliged.
We observed it as it approached the Sun over the eastern limb, disappeared
behind the Sun, and reappeared on the other side where it survived only
briefly before it vanished. We were entirely ready for it this time, even
performing a very rare set of off-point maneuvers with \textup{SDO}. I~realized
that this time, the observations provided a unique view of the solar corona
between the EUV-bright regions and the inner edge to where coronagraphs
operated: seeing the ionized material trace trajectories that outlined
the Sun's magnetic field provided a rare (and so far unique) opportunity
to validate global MHD models in 3D (Figure~\ref{fig:Lovejoy}). It took a
while to convince Jon Linker at Predictive Science Inc. of the promise
of this project, and of the likelihood of interesting the journal Science
in its publication, but a subcontract was set up and the results can be
found in the study by \cite{lovejoymhd2013} published a year and a half
later.

%

\subsection{The Largest Flares}
\label{sec7.5}

Another unanticipated opportunity to learn about solar activity, albeit
on the topic of statistics, came from the \textup{Kepler} spacecraft. Its
results triggered J\"urg Beer (who uses radionuclides to extract information
from ice cores, tree rings, and rocks) and me to propose an ISSI team
project: Kazunari Shibata had assembled a group of talented students to
search the \textup{Kepler} data for signatures of flaring on cool stars and
the first results suggested a power-law frequency distribution extending
that established for flares on the Sun to considerably more energetic events.
The International Space Science Institute is a marvelous environment for
what J\"urg and I had in mind, namely to bring together experts on solar-energetic
events and stellar flaring, along with experts in cosmogenic radionuclides
and their atmospheric transport and biospheric/geologic storage systems
to gather information on flare-energy distributions to assess the probability
of extreme flares on the Sun. Our team spent many hours assembled in Bern
discussing how to compare the disparate passbands in which flares were
observed, what kinds of signatures the associated energetic particles might
leave in terrestrial tree-ring and glacier records and in lunar rocks,
and how to bring all that information onto a common scale. One of the results
of the activities of that team was that a proposed connection between large
solar energetic-particle events and enhanced nitrate concentrations in
polar ice cores \citep{2001JGR...10621585M} was shown to be untenable.
At our first team meeting, Eric Wolff (a glaciologist and climate scientist)
walked over to the University of Bern to retrieve chemical analyses of
ice cores that showed that nitrate spikes were generally associated with
biomass-burning plumes, thus removing nitrate as a plausible tracer of
extreme flaring \citep{wolff+etal2012}. The ISSI team concluded that the
Sun likely could produce flares and particle storms of strengths not observed
within the half-century of flare monitoring. Follow-up analyses of the
\textup{Kepler} stars by Shibata and students left us with a less clearcut
conclusion: many of the superflaring stars turned out to be binary stars
that muddied the waters, leaving us uncertain about the most extreme solar
events \citep{2022LRSP...19....2C}.

\begin{figure*}
  \centerline{\includegraphics[width=0.75\textwidth,trim=0mm 0mm 0mm 50mm,clip=]{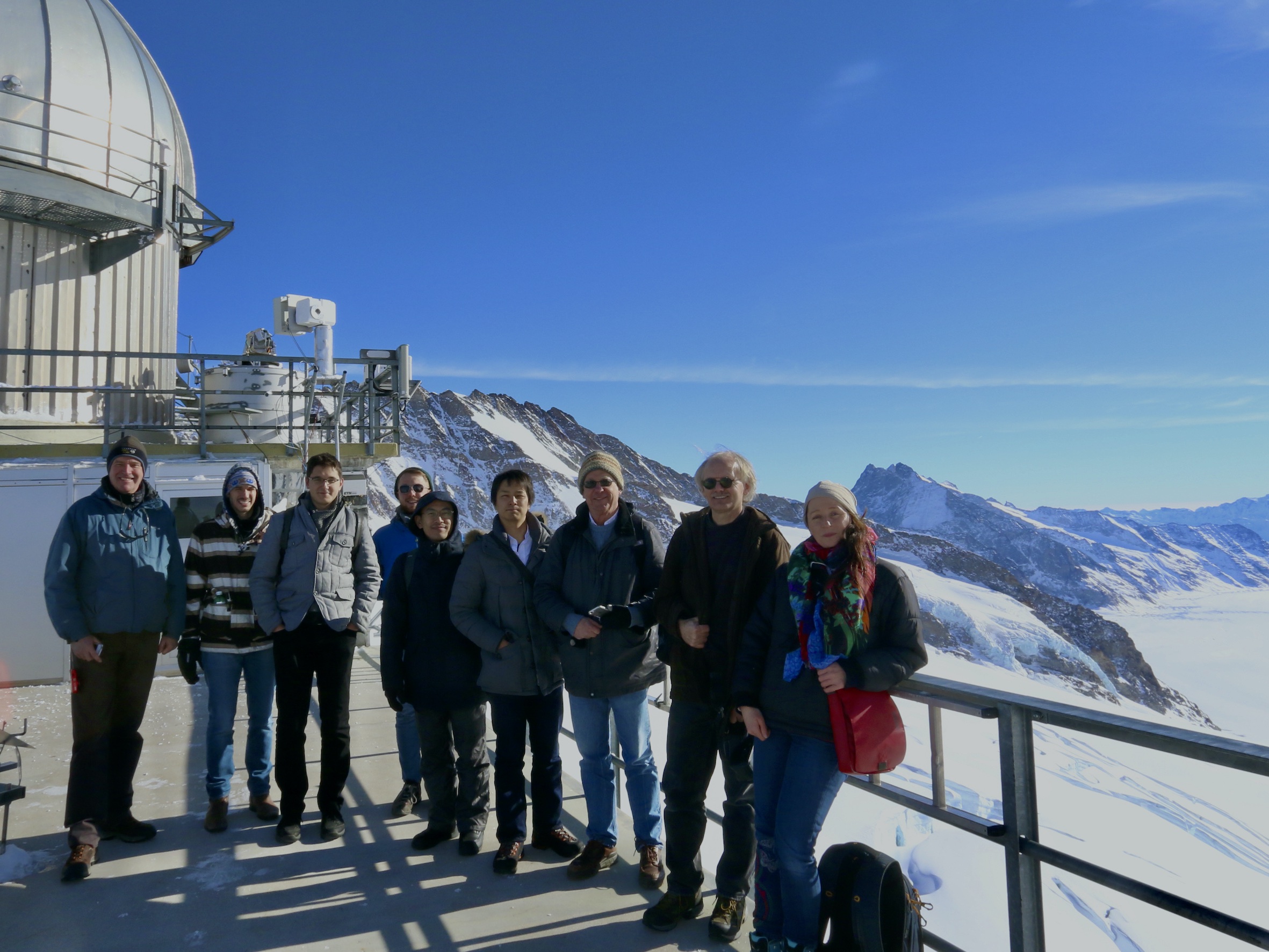}
              }
 \caption{I have always found the International Space Science Institute's
(ISSI's) welcoming environment and the many opportunities for relaxed,
informal discussions (where it is okay to show one's unfamiliarity with
the specialties of others) an efficient catalyst for discovery and learning.
This photo shows members of the ISSI team on ``Energy transformation in
solar and stellar flares'' (proposed by Louise Harra and Manuel G\"udel)
during their excursion to the Jungfraujoch Observatory. From left to right:
Karel Schrijver, Adam Kowalski, our observatory host, Magnus Woods, Hirohisa
Hara, Shin Toriumi, Hugh Hudson, Manuel G\"udel, and Theresa L\"uftinger
(missing are Louise Harra, Miho Janvier, Kanya Kusano, Sarah Matthews,
and Rachel Osten).}
\label{fig:HarraTeam}
\end{figure*}

A subsequent ISSI team project led by Louise Harra (Figure~\ref{fig:HarraTeam})
developed the rationale to look for the stellar equivalent of CMEs in coronal
dimmings \citep{issiharra}, since pursued in several theoretical and observational
studies
\citep[{e.g.},][]{2020IAUS..354..426J,2021NatAs...5..697V}.

\subsection{\textup{SDO} Beyond Our Peers}
\label{sec7.6}

\begin{figure*}
   \centerline{\includegraphics[width=0.75\textwidth,clip=]{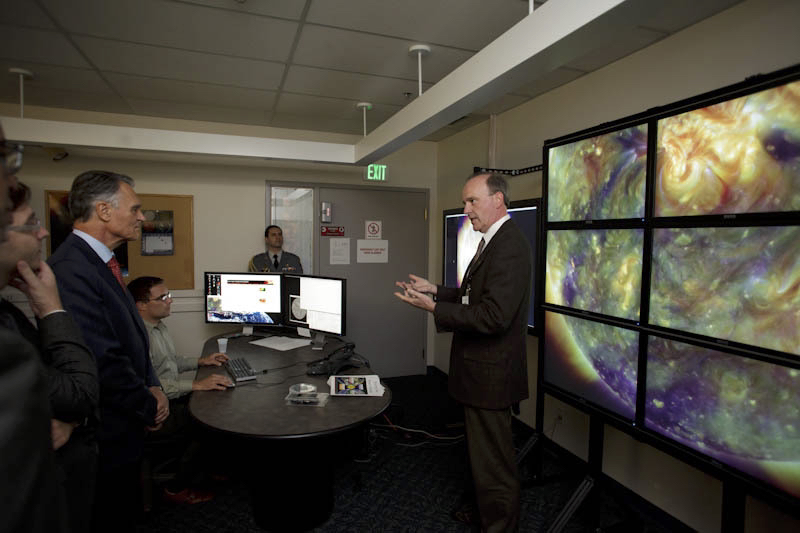}
              }
 \caption{The LMSAL video wall was used daily for the annotation of events
seen in the \textup{SDO}/AIA data as input for the Heliophysics Events Knowledgebase
(HEK) and to show the solar corona to visiting scientists and VIPs. Here,
I explain the purpose of our studies to An{\'\i}bal Cavaco Silva, then
president of Portugal, while Ralph Seguin (seated), who developed the display
system, drives the imagery.}
\label{fig:VideoWall}
\end{figure*}

\begin{figure}
  \centerline{\includegraphics[width=0.5\textwidth,trim=0mm 60mm 0mm 5mm,clip=]{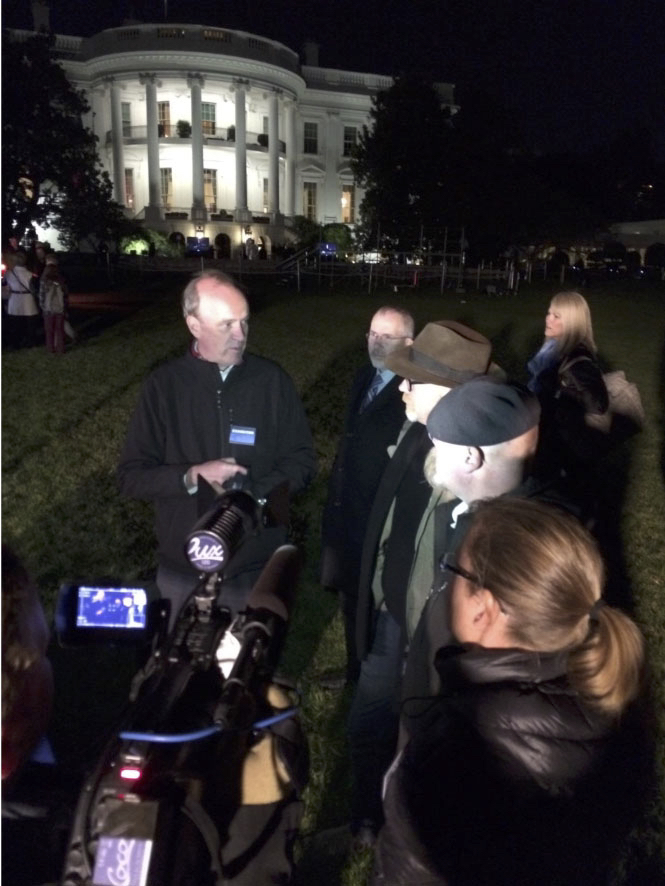}
              }
\caption{The second ``White House Astronomy Night'' hosted by President
Obama on 2015/10/19. Alison Nordt (bottom right, who later became LMSAL's
manager) and I brought a large 4K screen onto the White House lawn to show
\textup{SDO} images along with a model of the \textup{IRIS} spacecraft. Here
we are on either side of the Mythbusters (Adam Savage and Jamie Hyneman)
and staff of the Office of Management and Budget (OMB) explaining why solar
physics is important to society.}
\label{fig:astronomynight}
\end{figure}

It proved to be great fun to show the capabilities of AIA to solar physicists
and to our stellar colleagues (whose first reaction, upon seeing the early
imagery at a Cool Stars meeting, included mostly fascination but also a
discouraged ``I'll just have to give up and retire after seeing this'')
but also, of course, the general public (Figure~\ref{fig:VideoWall}). Within
a few years, AIA images made the covers of over 50 magazines ranging from
Nature to Sky and Telescope. We showed the imagery at public talks in museums
(including a near-live exhibit at the National Air and Space Museum developed
by our Harvard team members), in planetaria, and at amateur astronomy clubs,
domestic and abroad. I~even found myself invited to the White House for
an ``Astronomy Night'' on its lawn (Figure~\ref{fig:astronomynight}). Also,
to this day, an automated Google search of news media sends me an email
every day or two on \textup{SDO} science or images being included somewhere
in the world.

%

\section{Heliophysics Education}
\label{sec:heliophysics}

Next to doing research, a good part of my time was devoted to committee
work as a representative of our community and to the education of the next
generation of researchers. Little did I realize that in doing so, I would
embark on over two decades of educating myself. It started one morning
when George Withbroe, then Director of NASA's Sun-Earth Connection (SEC)
Division called me with the invitation to join the 2000 SEC Roadmap team
charged with developing a strategic plan for the next two decades. When
a few months later I found myself at NASA/HQ in DC surrounded by two dozen
of my peers, I could not believe that I was there. I~think that the formal
term is ``imposter syndrome.'' It is something that I experienced on each
of the twenty-some committees I have served on (including 25 years on the
editorial board of Solar Physics) amidst scientifically knowledgeable and
politically experienced peers (although the sense that they were by far
my seniors has faded over the years).

That Roadmap exercise made me realize that I had much to learn about not
only the neighboring sciences of heliospheric, magnetospheric, and Earth-atmospheric
physics, but also about solar physics itself. That Roadmap was followed
by an invitation to join the inaugural advisory committee for the Living
With a Star (LWS) program the year after. Of course, I was there to represent
the interests of solar physics, but I was also committed to advancing the
intimately connected sister sciences. That attitude at times raised eyebrows. I~remember, for example, a committee meeting in which I asked for a clarification
into arguments used in a discussion between ionospheric physicists, upon
which one of them asked me ``What's it to you? You're a solar physicist!''
That was a learning moment for me in an entirely different dimension. Although
the LWS program has enabled many more interfaces between its SEC (now Heliophysics)
subdisciplines, I fear that we have a long way to go before heliophysics
is viewed as a coherent science rather than as a collection of its subdisciplines.

I also had much to learn to recognize and navigate the politics that hides
below the surface in committees, including appreciating the importance
of choosing words with care. Using (what appears to be) the wrong word
can get you into trouble. Dick Fisher, the NASA Division director under
whom SEC became Heliophysics, experienced this when he proposed the new
name for the Division: there was a lot of criticism because it sounded
too much like solar physics to the majority of the scientists working in
the Division's fields. I~experienced his predecessor's, George Withbroe's,
unexpectedly angry retort when I suggested in a Roadmap meeting at NASA/HQ
that a mission to monitor the entire Sun in a variety of wavelengths would
be of great value to SEC: ``We do science at NASA! We do not `monitor'!''
Being a relatively junior member of the committee, I did not have the courage
to argue with him at that time that maybe I should have used the word observe,
but such a mission, \textup{SDO}, did, in time, emerge and its `monitoring'
led to many discoveries.

After learning a lot more during one more Roadmap and a 3-year membership
of the Space Studies Board of the National Academies of Sciences, NASA/HQ's
Madhulika Guhathakurta (LWS program scientist at the time) and I sat
down for lunch in a break at a meeting on ``Earth-Sun System Exploration''
in Hawaii to discuss a summer school in heliophysics. She brought me in
touch with the incredibly sage and patient George Siscoe (a professor
of space physics) with whom I submitted a proposal to NASA under the umbrella
of UCAR's Cooperative Programs for the Advancement of Earth System Science
(CPAESS) where the summer school has been ever since.\footnote{\url{https://heliophysics.ucar.edu/summer-school}.}

\begin{figure*}
  \centerline{\includegraphics[width=0.56\textwidth,clip=]{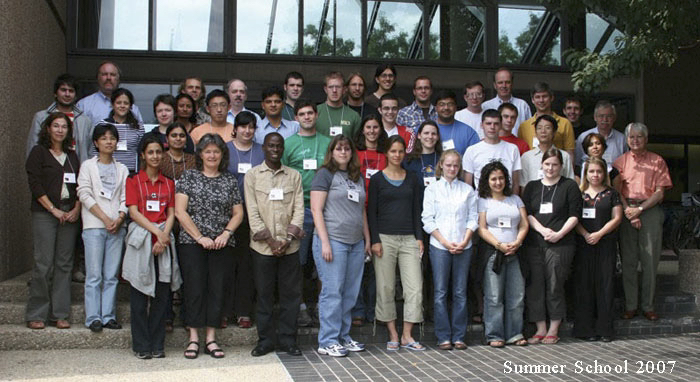}
                      \includegraphics[width=0.44\textwidth,clip=]{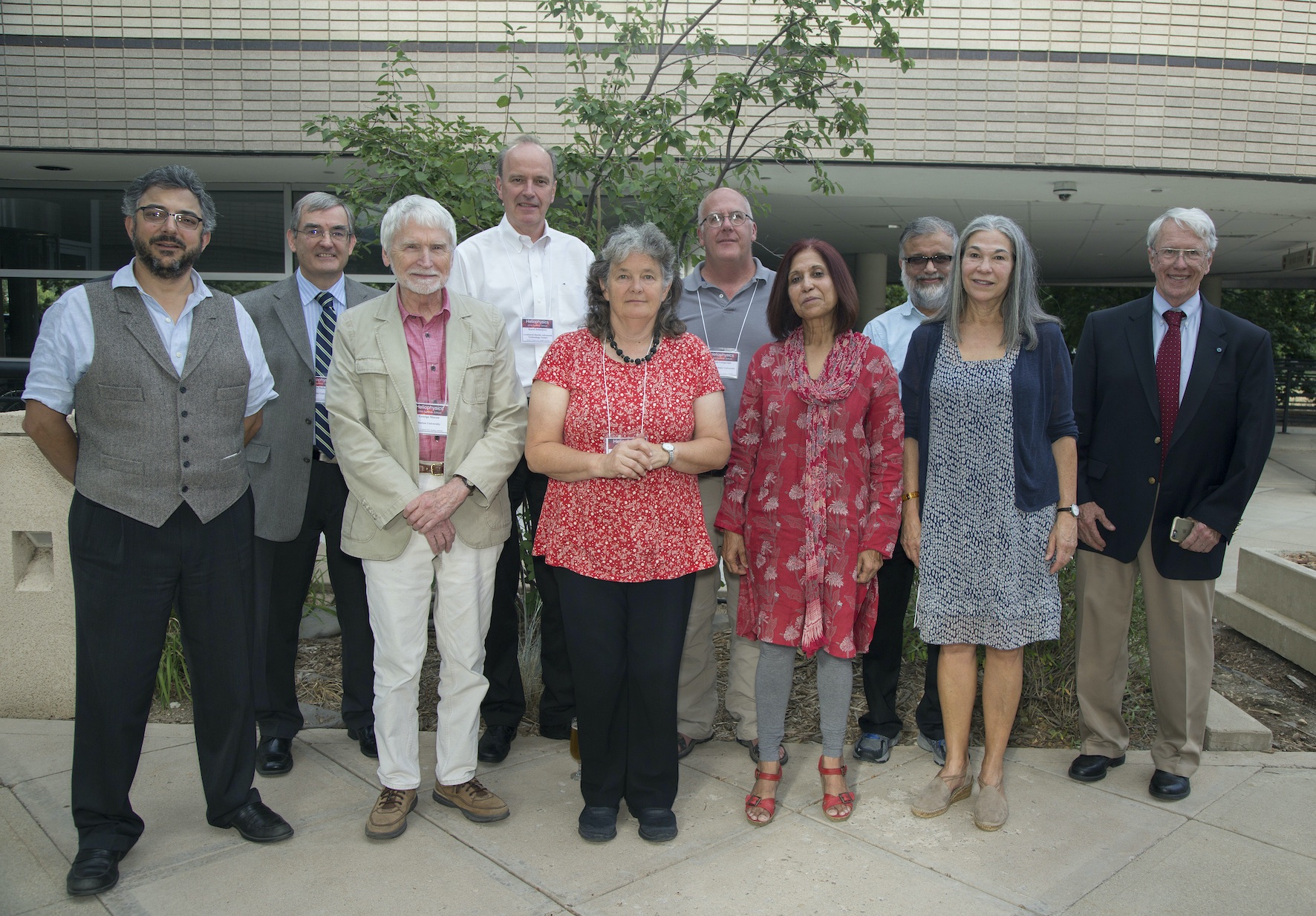}             }
 \caption{The first cohort of the Heliophysics Summer School in front of
the HAO building in Boulder, CO (\emph{left}), and the leadership of the
first decade of the school (\emph{right}): Nick Gross, Jan Sojka, George
Siscoe, Karel Schrijver, Fran Bagenal, Dana Longcope, Lika Guhathakurta,
Amitava Bhattacharjee, Meg Austen, and Dick Fisher (director of NASA's
Sun-Earth Connections Division at the time of the School's inception).}
\label{fig:HSS}
\end{figure*}

George's advocacy for ``universal processes'' and my interest in the sister
sciences of Heliophysics shaped the first Summer Schools held in 2007\,--\,2009
in Boulder, CO (Figure~\ref{fig:HSS}). I~learned so much from the talented
teachers that we had the privilege of working with, and they graciously
accepted the detailed editing of the written versions of their lectures.
However, boy, did I underestimate how much work editing three 400$+$-page
volumes of the Heliophysics series would be! Although George and I were
hoping for online books, NASA wanted printed volumes; we found Cambridge
University Press to be an efficient partner in this process (and benefited
greatly from the CPEASS staff who helped us in obtaining permissions to
use the many figures).

The largest challenge --~which persisted throughout the now 17-year history
of the School~-- was to get teachers to reach beyond their own discipline.
The challenge there is not only the uncomfortable unfamiliarity with neighboring
disciplines (demonstrated by the fact that George and I very rarely found
a given reference to appear in more than a single topical chapter), but
also the different uses of terms and even (apparently) fundamentally different
frames of thinking about things. Take, for example, the use of the word
dynamo, which can mean ``work applied against the magnetic field'' or ``the
process that maintains a magnetic field against decay,'' or the word convection,
which can mean ``advection'' or ``the hydrodynamic energy transport by
overturning motions.'' Another example is that ionospheric physicists think
in terms of electric field, ${\mathbf{E}}$, and current, ${\mathbf{j}}$, while their
solar-heliospheric colleagues use magnetic field, ${\mathbf{B}}$, and velocity,
${\mathbf{v}}$. Gene Parker expressed his unambiguous preference for
${\mathbf{B}}$ and ${\mathbf{v}}$ to avoid ``intractable global integro-differential
equations''; he notes that ``It is forgotten that the equations of Newton,
Maxwell, and Lorentz are self consistent. A~complete solution in terms
of ${\mathbf{B}}$ and ${\mathbf{v}}$ automatically implies the proper treatment of
${\mathbf{E}}$ and ${\mathbf{j}}$'' \citep{2014RAA....14....1P}.

That challenge of teaching beyond the stovepipe separation of subdisciplines
remained my main motivator for continuing to support the school. We were
not only educating the next generation of researchers (with some 600 students
participating thus far) but also giving our established peers views of
the sciences that they might otherwise not have time for. In the end, George
and I cowrote some of the chapters and provided endless (and likely annoying)
streams of hints to the lecturers/authors to reach across discipline boundaries.

After a three-year break, I engaged with the School for another cycle,
resulting in a fourth book, a volume that emphasized comparative heliophysics
of planets and stars surrounded by their astrospheres. That was far from
the end of this project, though. After I retired and had more time to pursue
things outside my direct professional commitments, I was awarded the 2018
Johannes Geiss Fellowship of ISSI. Spread out over two years, I spent several
months in ISSI's wonderful setting in Bern (and added some nice long hiking
vacations through the high Alps) creating the book that I --~and I think
also George Siscoe~-- had hoped for when we set out on the Summer School
project but that I lacked the knowledge to create at that time: with a
focus on ``universal processes,'' the resulting book, ``Principles of Heliophysics,''
presents a general introduction to heliophysics.

I also added 200 ``activities'' (homework exercises, discussion topics,
background reading, etc.), a feature we overlooked in the original books.
Only in 2021\,--\,2022 did I add ``solutions'' to the more challenging of these
exercises at the urging of Nick Gross who continues to be an indefatigably
upbeat partner in the School and without whom the practical ``lab'' side
of the School could never have come about. This final product is now available
at no cost on ar$\chi $iv and at printing cost on Amazon
\citep{2019arXiv191014022S}.

One envisioned volume was never completed. George and I hoped that a volume
dedicated to the mechanisms and impacts of space weather would be added
to the series. It was something that we hoped would grow with two or three
lectures/chapters each year. With changing leaderships and evolving focus
of the School, a completed, printed version never came to be, although
a beginning is available online at UCAR's CPEASS\footnote{\url{https://heliophysics.ucar.edu/sites/default/files/heliophysics/documents/HSS5.pdf}.}
(with apologies to its authors who were told a printed book would someday
appear).

Although this summer school provides an exploratory introduction to the
field of heliophysics, it is obviously impossible to transfer sufficient
knowledge in a 10-day experience to create anything approaching heliophysicists.
However, there is no place where heliophysics is taught as a discipline.
Heliophysics seems to be where the medical field was centuries ago, namely
in a master--apprentice setting rather than in its modern-day university
system of generalist-to-specialist training programs. Do we really expect
that broadly knowledgeable heliophysicists somehow materialize through
on-the-job training? With the rapidly growing field of comparative heliophysics,
i.e., the study of star-planet interactions in the broadest sense, is it
not high time to significantly invest in the education of the next generations
of researchers?

The broad perspective provided by heliophysics sparked an entirely different
project. As my wife and I discussed our professional activities, we realized
that although she came at life from a molecular geneticist's perspective
and I from the astrophysical side, there was an overlap when looking at
the transient nature of our existence: the body of every living thing on
Earth is in constant flux, with cells dying and being replaced, all with
materials made sometime in the Universe's history. We decided to write
a popular science book about all that entitled ``Living with the Stars''
(Oxford University Press). Iris and I enjoyed giving talks about this in
planetaria, in science museums, and to college students, sometimes even
together, emphasizing not only our transient nature but also that the idea
that we exist separate from nature is completely flawed: we are part of
the network of life and messing with our environment (from biodiversity
to climate) jeopardizes our very existence.

\section{Space Weather and Its Societal Impacts}
\label{sec:swx}

Space weather (SWx), by several other names, has a long history, advancing
from aurorae, geomagnetic variability, radio interference, and radiation
hazards for astronauts, to societal impacts through our growing technological
infrastructures on the ground and in space. The science of space weather
is recognized as a core focus of heliophysics, and the activities of the
solar-physics community aimed at understanding solar variability on time
scales from minutes to millennia are part of that. However, as I sat in
committee meeting after committee meeting, it began to bother me that investing
in the forecasting of space weather was given a high priority even though
SWx impacts seemed to be either minor for known events or speculative for
hypothetical extreme events. Studies were being published warning the world
of major, long-lasting power outages, of significant loss of satellite
communications and navigation systems, of particle radiation causing upsets
in aircraft avionics systems {\ldots} but apparently without an adequate
quantitative foundation of the likelihood or frequency of even the most
impactful events. I~add the qualifier ``apparently'' here because some
(in fact likely much) of that quantitative foundation is shielded from
the view of academic researchers, being considered competition sensitive
or a matter of national security. Nonetheless, not seeing that foundation
prompted me to attempt some assessments of impacts based on relatively
common space weather.

This became a possibility when Sarah Mitchell (my \textup{SDO}/AIA program
manager at the time) came across a website of the North American Electric
Reliability Corporation (NERC) where power-grid ``system disturbances''
were compiled. Comparing the timing of these events with geomagnetic records
from around the USA, we could establish that during times of enhanced geomagnetic
variability, the frequency of these system disturbances increased by a
small (4\%) but significant ($>$\,3$\sigma $) fraction. The publication of
that result put me on the invitation list for a space-weather meeting in
the catacombs of the US Capitol Building, where I was approached by Buddy
Dobbins from Zurich Risk Engineering. We worked out a way that information
on insurance claims to Zurich in the USA relating to electrical systems
could be used in a comparison with geomagnetic activity without running
into issues with client confidentiality. We concluded that regular space
weather has impacts through the electrical power grids in Europe and the
USA of the order of multiple billions of dollars each year, warranting
the research investment even without considering potentially devastating
extreme events \citep{2014SpWea..12..487S}.

Hoping to better quantify impacts, I hired an economics postdoc, supported
by NASA/Ames, to try to come up with financial equivalents of impacts.
It proved too hard, so we regrouped and decided to pursue a survey of the
customer interests and needs of the forecasts issued by NOAA's Space Weather
Prediction Center (SWPC). The SWPC leadership helped us by pointing its
user base to our online survey, resulting in almost 3000 responses. We
learned that SWx ``information is primarily used for situational awareness,
as an aid to understand anomalies, to avoid impacts on current and near-future
operations by implementing mitigating strategies, and to prepare for potential
near-future impacts that might occur in conjunction with contingencies
that include electric power outages or GPS perturbations''~\citep{schrijverrabanal2013}.
The users of these forecasts could not help us quantify the financial risks
of space weather.

Despite the improvement of my understanding of space weather through these
studies, the heliophysics summer school, and committee work, I was still
frustrated at what I saw in the political arena. Space-weather agencies,
both in the EU and USA, would formulate instrument requirements based largely
on desires for increased coverage by known types of instruments, albeit
perhaps from new vantage points, such as in an orbital position trailing
Earth to provide a side view of CMEs. This might work if the aim were,
for example, to improve the forecasts of the arrival time at Earth of CMEs
but it would not be of much help to the primary users of such information,
namely the power-grid and power-plant operators. This was made explicit
in one of many SWx meetings when, after the director of the SWPC claimed
that the uncertainty of the arrival window of CMEs could be substantially
reduced, I asked the NERC representative what he would do with such information.
He stated that he would not recommend their ``members'' in the power sector
take preventive action simply because the magnitude of the impact on the
power grid would still be unknown. After all, apart from arrival time,
we need to provide the magnetic-field orientation of the CME complex in
order to work through a model of the response of the Earth's magnetic field
(which itself is an enormous challenge). I've likened it in discussions
to telling people who are preparing for an outdoor wedding that the weather
might change in the course of the event without information about what
would change and by how much.

Of course, I was certainly not the only one troubled by the allocation
of significant resources with insufficient scientific knowledge. In a discussion
with Roger Bonnet and Madhulika Guhathakurta at the COSPAR meeting in Mysore,
India, in 2012, we came up with the idea of a COSPAR-sponsored strategic
roadmap on the science needs to improve SWx forecasts. As the saying goes,
be careful what you wish for: soon after, I was invited to chair that roadmap
exercise together with Kirsti Kauristie, an ionospheric-magnetospheric
scientist at the Finnish Meteorological Institute. COSPAR provided us
with the team and we set about with meetings, in person and by phone (Zoom
and Skype weren't accepted as tools by all the parties involved, most annoyingly
by my own organization), to shape a roadmap for the science of space weather
\citep{cosparrm2015}. It was a wonderful team: in a truly collaborative
spirit, we developed a consensus plan, starting with user requirements,
via science needs, to potential instrumentation. The roadmap was presented
at the 2014 COSPAR General Assembly in Moscow, Russia. Subsequently, Kirsti
and I presented it to various stakeholder agencies. We hope that the case
we made was heard: there is much to learn about the physics of space weather
before forecasts become reliably actionable.

\section{Retired?}
\label{sec:time}

\begin{figure*}
  \centerline{\includegraphics[width=0.43\textwidth,clip=]{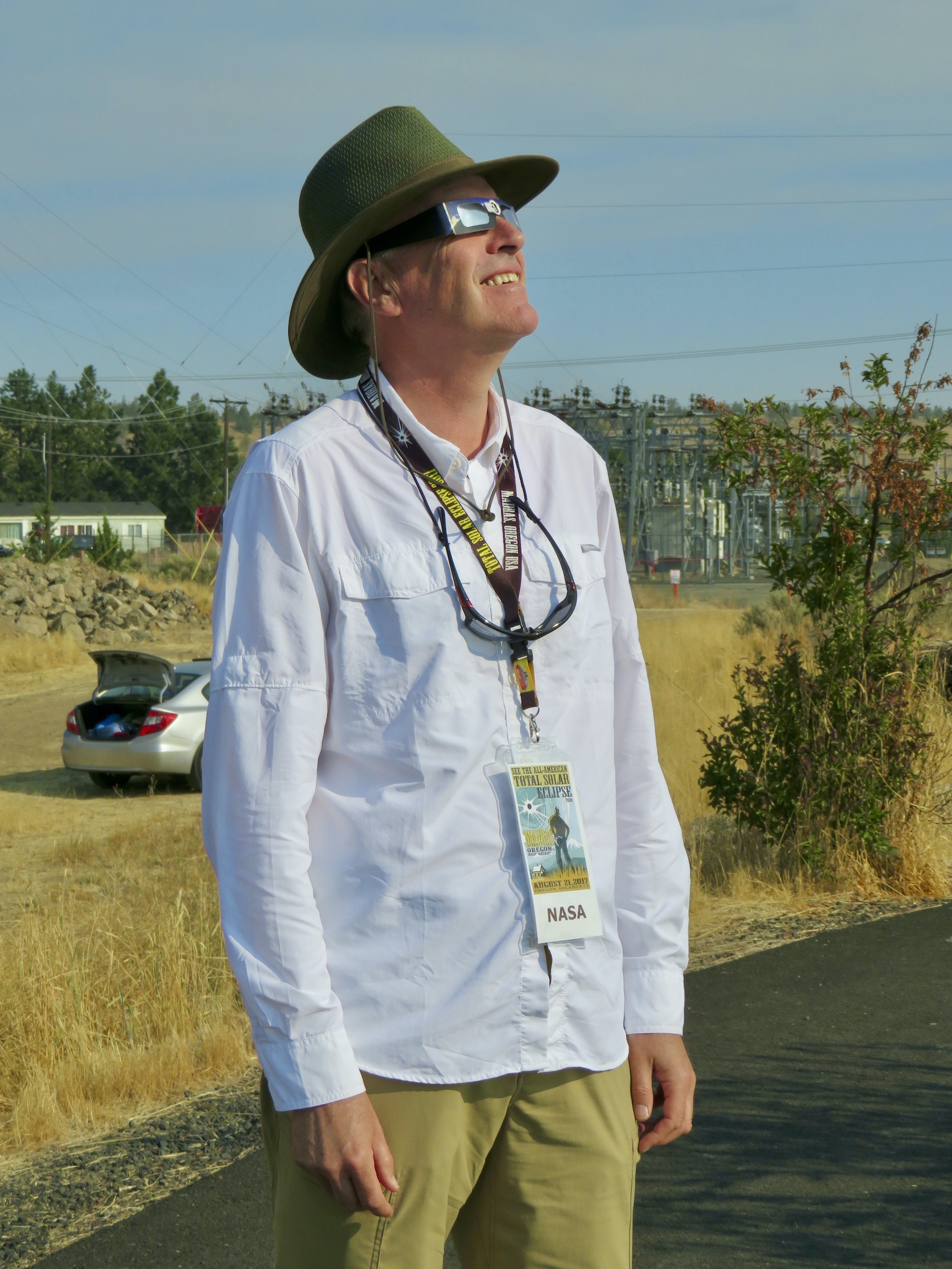}
              \includegraphics[width=0.57\textwidth,clip=]{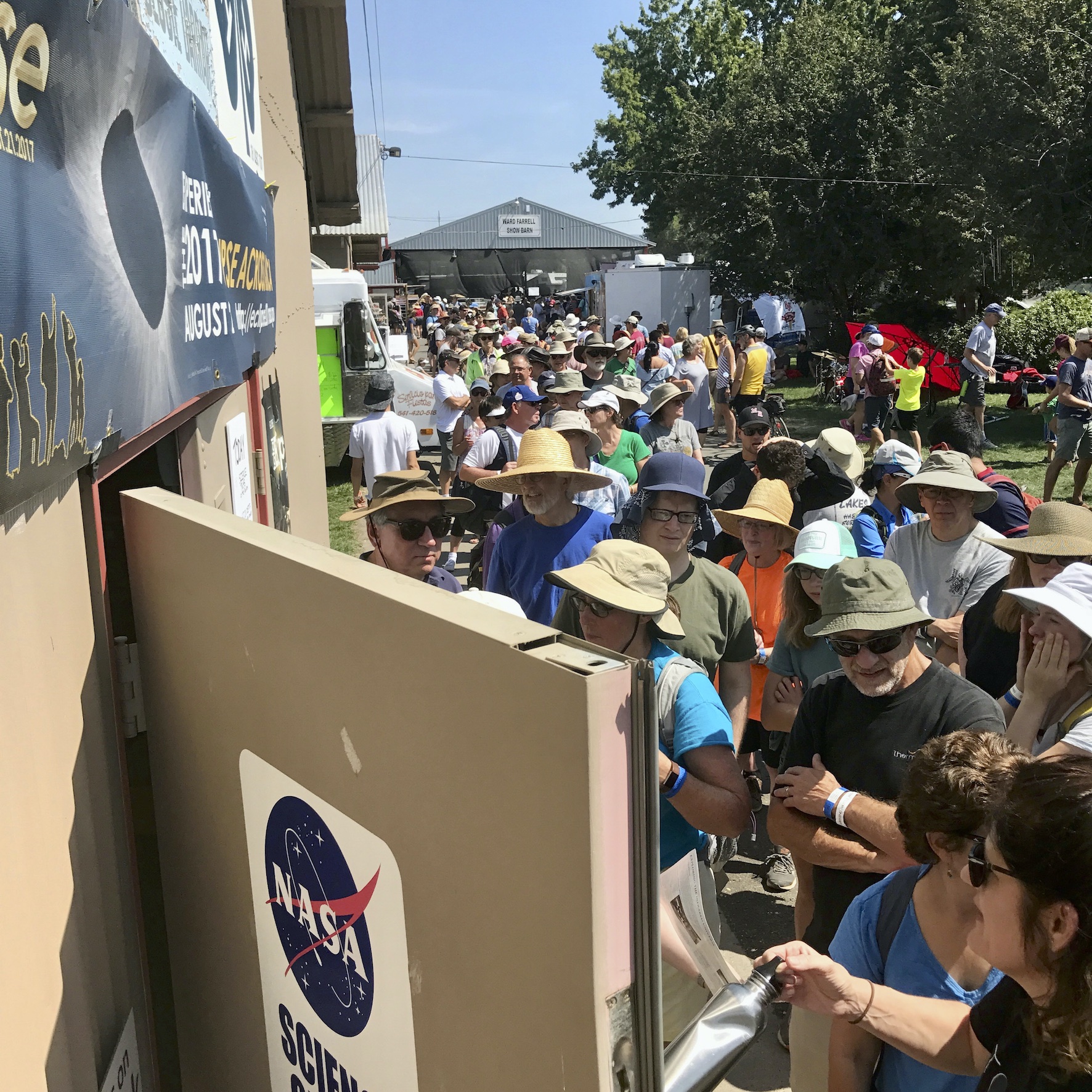}
                    }
\caption{The total solar eclipse of 2017/08/21 was the first such event
that I witnessed in person (\emph{left}), along with some 40{,}000 others
who flooded the small town of Madras, Oregon, and (\emph{right}) the barn
in which my NASA colleagues set up shop. It is the only venue that I have
ever been to where there was more interest in the presentations by the
scientists than in the live music being performed on the stage in the far
reaches of the county fairgrounds.}
\label{fig:Madras}
\end{figure*}

Life is short and uncertain, so my wife and I decided to step away from
employment fairly early. However --~yes, the principal of my grade school
was right~-- I have many interests, and science stayed with me. I~already
mentioned the book ``Principles of Heliophysics,'' the Heliophysics Summer
School, and work on exoplanetary transits, but how could I not be intrigued
by the discoveries about exoplanets, their stars, their planetary systems,
the issue of habitability, {\ldots}? Learning best by hands-on activities,
I wrote a book about it all entitled ``One of ten billion Earths: How we
Learn about our Planet's Past and Future from Distant Exoplanets''
\citep{2018otbe.book.....S}. Oh, and I finally gave in to seeing a total
solar eclipse with the naked eye. How could I not since, having moved to
Oregon, it came practically through our backyard (Figure~\ref{fig:Madras})?

%

\section{Good Fortune and a Flexible Frame of Mind}
\label{sec:lessons}

I have been incredibly lucky in my career, with remarkably fortuitous timing
relative to fascinating space missions, the good fortune to be employed
where the action was (it helped that I was very flexible in what I thought
of as ``the action''), and with amazing mentors, most notably my father
with his technical skills, Kees Zwaan with his big-picture views, Rolf
Mewe who reminded me to keep track of the details, Jeff Linsky who introduced
me to the world of proposals and grants and who brought me to the USA,
and Alan Title who introduced me to the politics of science and with whom,
for over two decades, I sat down most mornings before the duties of the
day for an often hour-long discussion of solar movies that he had just
made, of slide sets for presentations, of papers, of proposals, of instrument
designs, or of life in general. I~am grateful to them and to all others
who helped shape my path through science. I~am also grateful for the many
friendships that developed over the years.

In the Introduction, I promised a summing-up of critical observations
about the research enterprise of solar physics:
\begin{itemize}
\item \emph{Overcompartmentalization.} The solar and stellar branches of
the study of cool-star magnetic activity are pursued by essentially disjoint
communities and supported by segregated funding streams, as are the subdisciplines
of heliophysics. In the USA, ground- and space-based solar and stellar
research are supported by different agencies that focus more on integrating
hardware than knowledge. Agency-level international coordination and collaboration
is weak. Research funding is generally tied to specific observational platforms,
which makes it a high-risk --~and thus generally avoided~-- endeavor to
try to obtain coordinated observations (such as across the electromagnetic
spectrum) through separate observing proposals for existing observatories,
and even more so for proposed future ones. All this divides what could
be a coherent community, making colleagues into competitors.
\item \emph{A weak academic foundation.} Solar physics is not taught as a
science field in its own right, let alone the science of heliophysics (which
to me encompasses cool-star physics as well as the study of (exo-)planetary
systems). A~broad education prepares one to spot opportunities, readies
one to move with the trends of the day, and fosters appreciation of the
work of others in neighboring disciplines. In order to appeal to modern-day
students, we need to present them with an exciting, vibrant field full
of opportunities.
\item \emph{Piecemeal funding.} Too many of our colleagues --~certainly in
the USA~-- make do with small grants and contracts that support them for
at best months at a time. I~do not understand why this is: a desire to
micromanage, fear of the failure of larger projects, or community pressure
to fund many proposals, however inadequately, on the side of funding agencies,
or a community that finds fewer but larger grants too much of a continuity
risk in a highly competitive environment? Whatever the reason(s), one should
not expect frequent major breakthroughs from a few months of work. We are
partly at fault ourselves: we take it that teaming arrangements increase
chances of success for those few larger opportunities that exist (if they
are not required by the funders), but in doing so we effectively divide
such grants into small, less impactful parcels.
\item \emph{High cost of space missions.} The high cost of building and launching
space missions translates into infrequent mission opportunities, which
indirectly leads to peculiar outcomes from this scientist's point of view:
a \textup{STEREO} mission without magnetographs; vector-magnetographic instruments,
including that on \textup{SDO}, without a demonstrated ability to successfully
utilize the data; a \textup{Solar Orbiter} with high-resolution instruments
without the needed telemetry and an orbit that neither corotates with the
Sun nor reaches the orbital inclination initially listed as scientific
requirements; a Living With a Star program and a Space Weather Enterprise
in which the component missions meant to study the system as a whole do
not adequately cover it, sometimes not even overlapping in time
{\ldots} To improve on this state, (1) there need to be more opportunities
that are less costly to build and operate (and, in the case of spacecraft,
cheaper to launch), and (2) there needs to be increased international cooperation
led by the science community.
\end{itemize}
\par
When I review my life lessons as a researcher, these come out at the top:
\begin{itemize}
\item Never jeopardize your integrity.
\item Treat everyone with respect, and learn from their viewpoints.
\item Respect the time your audience gives you when you give a presentation:
treat it like a performance, well-practiced and -timed, just as actors,
musicians, or other stage performers would do.
\item No matter how ``difficult'' a referee or reviewer may seem, carefully
consider the advice: these colleagues generally mean to help you by pointing
out that something can be improved about your methods, analysis, or presentation.
\item Be willing to move into new research (and funding) areas, even though
you will feel ``out of your depth'' as you dive into them: discoveries
often lie at the interface of disciplines where ``fresh eyes'' may spot
them.
\item Be guided by intuition but work your logic hard.
\item Do not move away from a situation that makes you unhappy unless you
have worked out where you would like to find yourself instead.
\item If your desired next job slips by you, consider that, in the end,
you may find that new paths present themselves that are as rewarding.
\item Share your knowledge and your data: there is plenty to discover.
\item Collaborate with colleagues and support the community, both nationally
and internationally: working in teams, panels, and committees broadens
your expertise, gives you visibility, and offers new perspectives.
\item A career in science is challenging, employment is never certain,
and willingness to move around is required, but it is wonderfully rewarding
to discover things, spot connections, and deepen your understanding of
the Universe!
\end{itemize}

\acknowledgments{Many of my scientific colleagues who have helped me
  in my development as heliophysicist are mentioned above
  but there are many more who remained \rc{unnamed} to avoid this
  manuscript from becoming unreasonably long. There are
  many others still who helped somewhere along the way
  without whom things simply would not function or
  even exist: the hardware and IT engineers; those who make
  conferences, workshops, and summer schools actually happen;
  secretaries; administrators; the building support staff; and more. I
  thank you all. I thank Ed Cliver, \rb{Alan
  Title, Bert van den Oord, and the referee} for comments on this
  manuscript. \ra{In writing this manuscript (and many others over the
  past thirty-odd years) I have made extensive use of the
  research-sharing platform ar$\chi$iv and of NASA's Astrophysics Data
  System Bibliographic Services.}
} 


\bibliographystyle{spr-mp-sola}
\bibliography{../ref_karel}

\end{article} 

\end{document}